\documentclass[journal]{IEEEtran}
\usepackage{stfloats}

\usepackage{cite}
\usepackage{amsmath,amssymb,amsfonts,mathtools,bm,bbm}
\usepackage{graphicx}
\usepackage{textcomp}
\usepackage[usenames,dvipsnames]{xcolor}
\usepackage{mathbbol}
\usepackage{enumitem}
\setlist[itemize]{leftmargin=*}
\setlist[enumerate]{leftmargin=*}
\usepackage{soul}

\let\oldFootnote\footnote
\newcommand\nextToken\relax

\renewcommand\footnote[1]{%
    \oldFootnote{#1}\futurelet\nextToken\isFootnote}

\newcommand\isFootnote{%
    \ifx\footnote\nextToken\textsuperscript{,}\fi}

\usepackage{standalone}
\usepackage{pgfplots, pgfplotstable}
\pgfplotsset{compat=newest} 
\usetikzlibrary{pgfplots.groupplots}
\usepgfplotslibrary{colorbrewer}
\pgfplotsset{cycle list/Dark2}

\usepackage{hyperref}
\hypersetup{
    colorlinks=true,
    linkcolor=blue,
    filecolor=magenta,      
    urlcolor=cyan,
    pdfpagemode=FullScreen,
    citecolor=teal
    }
\usepackage{cleveref}
\def\BibTeX{{\rm B\kern-.05em{\sc i\kern-.025em b}\kern-.08em
  T\kern-.1667em\lower.7ex\hbox{E}\kern-.125emX}}
  
\usepackage{amsthm}
\newtheorem{thm}{Theorem}
\newtheorem{lemma}{Lemma}
\newtheorem{defn}{Definition}
\newtheorem{cor}{Corollary}
\newtheorem{remark}{Remark}
\crefname{lemma}{lemma}{lemmas}
\Crefname{lemma}{Lemma}{Lemmas}
\crefname{thm}{theorem}{theorems}
\Crefname{thm}{Theorem}{Theorems}
\crefname{remark}{remark}{remarks}
\Crefname{remark}{Remark}{Remarks}
\crefname{defn}{definition}{definitions}
\Crefname{defn}{Definition}{Definitions}
\crefname{cor}{corollary}{corollaries}
\Crefname{cor}{Corollary}{Corollaries}

\DeclareRobustCommand{\T}{\intercal}
\usepackage{algorithm}
\usepackage[]{algpseudocode}

\newcommand{\tr}[1]{\mathrm{tr}(#1)}

\newcommand{\qtilde}[0]{\tilde{\bm{q}}}

\newcommand{\XX}{\mbox{\tiny \it XX'}}

\allowdisplaybreaks

\usepackage{subfig}

\usepackage{nomencl}
\makenomenclature
\begin{document}
\title{Differentially Private $K$-means Clustering\\ Applied to Meter Data Analysis and Synthesis}
%
%
%

\author{Nikhil~Ravi,
        Anna~Scaglione,
        Sachin~Kadam,
        Reinhard~Gentz,\\
        Sean~Peisert,
        Brent~Lunghino,
        Emmanuel~Levijarvi,
        and~Aram~Shumavon
\thanks{N. Ravi and A. Scaglione are with the Department of Electrical and Computer Engineering, Cornell Tech, e-mail: nr337@cornell.edu.
S. Kadam is with Arizona State University.
R. Gentz did this work while employed at Lawrence Berkeley National Laboratory.
S. Peisert is with Lawrence Berkeley National Laboratory.
B. Lunghino, E. Levijarvi, and A. Shumavon are with Kevala, Inc.
The authors would also like to acknowledge Kaiya Levine.
}
\thanks{
This research was supported by the Director, Cybersecurity, Energy Security, and Emergency Response, Cybersecurity for Energy Delivery Systems program, of the U.S. Department of Energy, under contract DE-AC02-05CH11231.  Any opinions, findings, conclusions, or recommendations expressed in this material are those of the authors and do not necessarily reflect those of the sponsors of this work.}%
}

%


\maketitle

\begin{abstract}
The proliferation of smart meters has resulted in a large amount of data being generated. 
It is increasingly apparent that methods are required for allowing a variety of stakeholders to leverage  the data in a manner that preserves the privacy of the consumers. The sector is scrambling to define policies, such as the so called `15/15 rule', to respond to the need.  However, the current policies fail to adequately guarantee privacy. In this paper, we address the problem of allowing third parties to apply $K$-means clustering, obtaining customer labels and centroids for a set of load time series by applying the framework of differential privacy.
We leverage the method to design an algorithm that generates differentially private synthetic load data consistent with the labeled data. 
We test our algorithm's utility by answering summary statistics such as average daily load profiles for a 2-dimensional synthetic dataset and a real-world power load dataset.
\end{abstract}


%
\IEEEpeerreviewmaketitle

\printnomenclature

\nomenclature[02]{$X$}{The subset of the superset that is queried}
\nomenclature[03]{$P$}{The number of datapoints queried}
\nomenclature[04]{$\bm{x}_p$}{The $p$-th datapoint, e.g., power load of a house}
\nomenclature[05]{$d$}{The dimension of each data point}
\nomenclature[06]{$\mathbbm{q}(X)$}{The query}
\nomenclature[07]{$\bm{q}$}{The query answer}
\nomenclature[09]{$\tilde{\bm{q}}$}{The Differentially private query answer}
\nomenclature[10]{$\bm{c}_i$}{The centroid of cluster $i$}
\nomenclature[11]{$\ell_p$}{The label of point $p$}
\nomenclature[111]{$r(X,\bm{q})$}{The clustering loss}
\nomenclature[112]{$\bar{r}$}{The DP accuracy loss}
\nomenclature[12]{$\epsilon$}{The privacy budget}
\nomenclature[13]{$\epsilon_c$}{The privacy budget associated with the centroids}
\nomenclature[14]{$\epsilon_{\ell}$}{The privacy budget associated with the labels}
\nomenclature[15]{$\delta$}{The privacy guarantee}
\nomenclature[16]{$\delta_c$}{The privacy guarantee associated with the centroids}
\nomenclature[17]{$\delta_{\ell}$}{The privacy guarantee associated with the labels}
\nomenclature[18]{$\Delta\mathbb{c}$}{The sensitivity of the centroid query}
\nomenclature[19]{$\Delta\mathbb{l}$}{The sensitivity of the label query}

\section{Introduction}

\IEEEPARstart{T}{he} growth in capabilities for data collection and computation has led to better products and greater market efficiencies in many sectors \cite{diamantoulakis2015big,zhang2018big}. In this paper, we consider the case of electric utilities that, over the last decade, have significantly expanded their residential metering and sensor deployments over distribution feeders~
responding to regulation~
for demand forecasting~
 transactive energy, power flow optimization, fault-detection, planning for distributed energy resources (DERs), and for billing and automatic disconnection  (see e.g., \cite{wang2018review} for a taxonomy of the applications). An important commercial application for the data is to target consumers for promotional campaigns, mapping their behavior in classes ~
to improve pricing or to provide incentives for reducing or shifting consumption, or for installing DER. 
Among the data queries that are useful to analyze customer data, \textit{clustering} is one of the most common (see e.g., \cite{yang2013review} for a survey) because of its unsupervised nature and relatively high accuracy.
In \cite{ramos2012typical}, the authors present a variety of clustering techniques to identify typical {\it daily load profile} of consumers, and in \cite{sharma2014electrical}, the authors propose a $K$-means clustering technique to identify similar types of load profiles for demand variation analysis and energy loss estimation. 


None of the papers on clustering cited above addressed the issue of privacy in releasing the query results on customers' smart meters' database.  Safeguarding against unintended disclosure of private data is critical in this sector. In fact, since the early deployments of residential smart meters, many researchers have investigated methods to maintain privacy while still enabling the data to be useful (see \cite{asghar2017smart} for a survey).  The most conventional methods in the industry are access control (e.g., see~\cite{ruj2013decentralized}), ``anonymization'' (e.g., see~\cite{efthymiou2010smart}), and data falsification techniques. Access control techniques alone allow all or no access.  Anonymization masks data or makes it more general, but it has been shown repeatedly~\cite{Narayanan2014No-silver-bulle}, that it is either removing too much of the data so that the query becomes useless or that it is typically insufficient, enabling linkage attacks that can be used to re-identify records. For example, researchers were able to re-identify some users in the anonymized data consisting of AOL search engine queries even when user IDs and IP addresses were removed~\cite{barbaro2006face}, and similar approaches were used to re-identify anonymized records in the Netflix Prize dataset~\cite{Narayanan2008-short} and the Personal Genome Project~\cite{sweeney2013identifying}.  A further disadvantage of anonymization is that once the data set is released, it is forever vulnerable to future reidentification attacks.  For electric grid data, regulators have proposed several policies to share electric consumer data in the public domain. Of particular note is the “15/15 Rule''~\cite{15_15rule} which states that any aggregation of customer data is considered anonymous if it contains at least fifteen customers and if no single customer's data comprises $15\%$ or more of the total values in the aggregated answer. There is no scientific rationale behind this rule; in fact, we show in Section \ref{sec:1515} that it offers no privacy guarantee. 

With this in mind, this paper presents an approach to applying differential privacy (DP)~\cite{dwork2006calibrating} for releasing the clustering results of the $K$-means algorithm to allow several stake-holders  to query the meters' data while preserving privacy. We also use the mechanism as a stepping stone to publish differentially private synthetic data that emulate the consumer behavior in a class.  DP consists of a growing suite of randomized methods to publish the output of data queries while guaranteeing that even multiple query answers are statistically unlikely to reveal information about an individual's data contributions, or lack thereof.  Within the DP framework, there is no release of raw data, only the output of differentially private queries.  Because DP acts as a ``guard'' between the query and response, DP is also capable of ceasing to respond to future queries once a pre-defined ``privacy budget'' has been reached.  Another important aspect of DP methods is that they are tailored to the query and the database that is queried, which explains why our paper focuses on clustering. In contrast to anonymization techniques, DP mechanisms corrupt the query answers and thus, produce relatively lower accuracy, depending on the amount of noise added. However, the accuracy-privacy tradeoff is analytically quantifiable through DP.
%
%

Generic differentially private clustering techniques have been previously presented in \cite{yu2016outlier,balcan2017differentially,ren2017dplk,xia2020distributed,lu2020differentially,ni2021utility}. The authors in \cite{yu2016outlier} proposed a heuristic-based outlier-eliminated DP clustering mechanism with adaptive Laplacian noise. The authors in \cite{balcan2017differentially} proposed an iterative $K$-means clustering algorithm for data in high-dimensional Euclidean spaces. In \cite{xia2020distributed}, the authors proposed a local DP iterative clustering algorithm where noise is added at the user's end before transmitting the data to the aggregator. While guaranteeing DP, these techniques may not converge. Instead, the authors in \cite{lu2020differentially} proposed a clustering algorithm that performs an input perturbation in each iteration, which offers convergence guarantees but drives the cost of DP higher depending on the number of iteration required for convergence. 
The authors of \cite{ren2017dplk,ni2021utility} both propose methods with better initial point selection to improve the clustering accuracy, where they first generate $K' \gg K$ centroids by running the adaptive DP K-Means algorithm on randomly divided subsets of the dataset, before merging them into $K$ centroids through an iterative process.
These papers do not consider the privacy loss on the publication of cluster labels, which is ultimately what the classifier wants to know.  Also, we note that the data such as a house's load profile are relatively smooth, and the i.i.d. noise added can be partly filtered out, whereas our proposed randomized mechanism leverages this fact to improve the performance.  

Having classified the customers, synthetic load models that emulate the corresponding profiles are an important tool in power system studies to run realistic simulations. To overcome the shortage of publicly available large scale load datasets, researchers have tried to fill the gap by either publishing anonymized real data or creating synthetic datasets using historical datasets~\cite{pinceti2019data,el2020data}. Recently, researchers have adapted the use of conditional Generative Adversarial Networks (GAN) to generate realistic, synthetic week-long time-series load profiles at high resolutions~\cite{fekri2020generating, pinceti2021synthetic}. But these frameworks lack privacy guarantees for the historical data used in the process. To make matters worse, if the distribution of the generated synthetic data is close to the distribution of the true historic data, then any privacy leaked by the estimated statistics from the true data will also be leaked by the synthetically generated data.
Also, we argue that the statistical structure of each class data can be approximated well by a multivariate log-normal distribution,  whose generation is far less complex to train and use compared to a GAN. 

A synopsis of our contributions is as follows:
\begin{enumerate}[leftmargin=*]
    \item We propose a novel differentially private clustering mechanism, which to the best of our knowledge contains:
    \begin{itemize}
        \item The first analysis of the privacy leakage on the publication of the noisy labels and the first mechanism to publish them with DP guarantees.
        \item The first of its kind, optimum scheme for adding \textit{colored} Gaussian noise to the cluster centroids, that we recently proposed in \cite{ravi2021colored}. This mechanism also provides greater accuracy for a given privacy budget compared to adding white noise, and also performs better in terms of privacy leakage compared to existing literature.  
    \end{itemize}
    \item We also provide a mechanism that, leveraging the empirical good fit of Advanced Metering Infrastructure (AMI) data with a mixture of multivariate log-normal vectors, generates differentially private synthetic load data for each cluster, that can be safely published for simulation studies.
\end{enumerate}
Finally, we test the efficacy of the proposed mechanism on samples drawn from a Gaussian mixture and a real-world AMI dataset. We wish to remark that the publication mechanism proposed for labels and centroids is broadly applicable to any type of data, while the generation of synthetic data relies on the AMI statistics and, thus, is domain specific. 

{\bf Paper organization}: In \Cref{sec:prelims}, we introduce our problem statement, discuss the threat model, before introducing the DP framework. In \Cref{sec:additive}, we describe a DP mechanism for the publication of the clustering query and in \Cref{sec:use-case}, we present a model to generate synthetic load profiles. Finally, in \Cref{sec:numericals}, we numerically test our algorithms, before concluding this paper in \Cref{sec:conclusion}.

{\bf Notation:} Boldfaced lower-case (upper-case respectively) letters denote vectors (matrices respectively) and  $x_i$ ($X_{ij}$ respectively) denotes the $i$\textsuperscript{th} element of a vector $\bm{x}$ (the $ij$\textsuperscript{th} entry of a matrix $\bm{X}$ respectively). Calligraphic letters denote sets and $|\cdot|$ their cardinality. Finally, $[N]$ denotes the set of integers $\{0,1,\ldots,N-1\}$.

\section{Preliminaries}\label{sec:prelims}
In the following, we denote by $X$ a set of feature vectors $\bm x_p$, $p\in [P]$ embedded in $\mathbb{R}^d$ that are in a database $\mathcal{X}$, which we can organize as a $P\times d$ matrix $\bm{X} := [\bm{x}_1, \ldots, \bm{x}_P]^\T$.
Specifically, in an energy system that is considered in this paper, $\mathcal{X}$ is the set of homes that a utility company serves, $\bm{x}_p$ is the time-series of meter data of house $p$.
To review the basic concepts, we set the problem in general terms and denote by $\mathbb{q}(X)$ the function that maps $X$ onto the query answer, and denote the outcome by  $\bm{q} \in \mathcal{Q}$, where $\mathcal{Q}$ is the domain of the query answers.

\subsection{Is the ``15/15'' Rule sufficient?}\label{sec:1515}
For energy systems, the 15/15 rule focuses on an averaging query:
\begin{equation}
    \mathbb{q}(X) = (1/P)\textstyle{\sum_{x \in X}} x,
\end{equation}
where $P \geq 15$ and $\forall x\in X$, $x \leq 0.15~\mathbb{q}(X)$. 
The claim of the 15/15 rule is that if one abides by it, then the data points are anonymized. However, there is a fundamental flaw with this type of aggregation when it comes to the privacy of each of the data records $x \in X$. Let us now consider an adversary who has queried for the average of the dataset $X$ and  a neighboring dataset $X'$, which differ from  $X$ by just one point $x_0 \in X$, i.e., $X' = X \setminus \{x_0\}$. If both $\mathbb{q}(X)$ and $\mathbb{q}(X')$ are revealed to them, then by simple algebra:
\begin{equation}
    x_0 = P\mathbb{q}(X) - (P-1)\mathbb{q}(X').\label{eq:infer}
\end{equation}
Thus, they are able to infer the value of the point $x_0 \in X$ in the scenario described above, no matter what $P$ is. This shows the need for a randomized response to the aggregate query, because responding with the exact answer to repeated queries will lead to privacy leakage. Next, we introduce our problem setup. 

\subsection{Problem Statement and Threat Model} 

\begin{figure}[!htbp]
    \centering
    \includegraphics[width=0.45\textwidth]{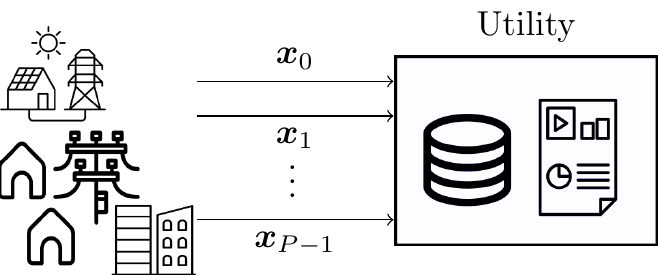}
    \caption{The utility collects the raw AMI data from the customers it serves.}
    \label{fig:data_collection}
\end{figure}

\begin{figure}[!htbp]
    \centering
    \includegraphics[width=0.45\textwidth]{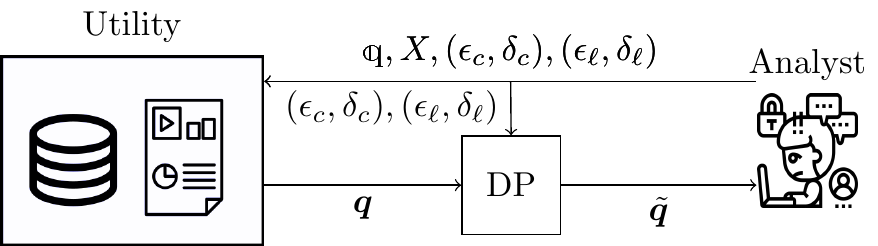}
    \caption{A third party analyst may query the dataset.}
    \label{fig:analyst}
\end{figure}

In this paper, we consider the scenario where a trusted central data owner collects and stores the data and untrustworthy third parties query them. For example, an electric utility collects and stores the AMI data of the customers as seen in \cref{fig:data_collection}. 
The electric utility agents, with legitimate access to the data, use them to perform operational functions such as scheduling, billing, etc., without being hindered by noisy data\footnote{There exists a second school of DP works, including \cite{lou2017cost}, called Local DP, where there is no central data owner with access to the raw data. Here, individual customers add local noise to their own data before forwarding it to the data owner. In such scenarios, there is a significant cost to the operational functions of the electric utility, as it now does not possess the raw data it needs. 
}. But the electric utility must abide by strict laws pertaining to customer privacy while publishing the data or aggregate query answers to external agents. Third party agents may query ($\mathbbm{q}$) the utility's data, and receive a differentially privatized answer ($\tilde{\bm{q}}$) instead of the true query response ($\bm{q}$), as shown in \cref{fig:analyst}. The quality of the DP response depends on the budget ($(\epsilon_c,\delta_c),(\epsilon_{\ell},\delta_{\ell})$) that the analyst is willing to spend to get the answer. The concept of privacy budget is explained in detail later in \Cref{sec:dp_prelims}. 

The goal of this paper is to study a differentially private randomized mechanism that would allow an analyst to perform a $K$-means clustering query on $X\subseteq \mathcal{X}$; that is, the analyst will obtain the information about the centroids as well as the labels of the data points in $X$ through a randomized algorithm that will make it either difficult or impossible to tell if the database $X$ or $X'$ was queried, with $X'$  missing one feature vector relative to $X$. The randomized mechanism proposed involves adding structured noise to the cluster centroids and to a subset of the labels. Prior to describing the mechanism, we review both 
the $K$-means clustering algorithm and the basic notion of DP next.

\subsubsection{Review of $K$-means clustering}\label{sec:kmeans_prelims}
The $K$-means algorithm splits the dataset, into $K>1$ subsets (clusters) and assigns a label to each point corresponding to the nearest cluster centroid to itself. In other words, the query we are interested in is given by $\mathbb{q}~:~\mathbb{R}^{P \times d}~\rightarrow~[K]^{P}~\times~\mathbb{R}^{K \times d}$. The problem can formally be posed as an optimization problem of the form:
\begin{subequations}\label{eq:clus}
\begin{align}
  \arg\min_{\mathfrak{C}} &~~ \frac{1}{P} \sum_{k \in [K]}\sum_{p \in \mathcal{C}_k} \|\bm{x}_p - \bm{c}_k\|_p\label{eq:clus_obj}\\
  \text{s.t.} &\quad \bigcap_{k \in [K]} \mathcal{C}_k = [P],\label{eq:clus_cons}
\end{align}
\end{subequations}
where $\mathfrak{C} = \{\mathcal{C}_1,\ldots,\mathcal{C}_K\}$ is the partition in $K$ clusters and $\bm{c}_k \in \mathbb{R}^d$ is the centroid of a cluster $\mathcal{C}_k$, obtained by averaging the points $\{\bm{x}_p\}_{p\in \mathcal{C}_k}$. The objective in \cref{eq:clus_obj} minimizes the cost of clustering assignment $\mathfrak{C}$ with the constraint in \cref{eq:clus_cons}, so  that every point in the database is assigned a cluster label whose centroid is the closest. 
The output of this algorithm consists of two components: 
\begin{equation}
  \mathbb{q}(X) := \{\mathbb{c}(X),\mathbb{l}(X)\},\label{eq:clustering_components}
\end{equation}
where
\begin{enumerate}[leftmargin=*]
    \item $\mathbb{c}(X)$ -- with outcome $\bm{c} = [\bm{c}_0^\T,\ldots,\bm{c}_{K-1}^\T]^\T$ is the sub-query that returns the set of $K$ centroids $\{\bm{c}_k~\in~ \mathbb{R}^d\}_{k\in[K]}$. 
    \item $\mathbb{l}(X)$ -- with outcome $\bm{\ell}$ is the sub-query that assigns a label to each feature point $\bm{x}_p$ that corresponds to an index of such centroids. 
\end{enumerate}
The performance of the $K$-means query outcome, $\bm{q} = \{\bm{c},\bm{\ell}\}$, is measured in terms of the \textbf{\textit{clustering loss}} which is defined as:
\begin{defn}[Clustering Loss]
    The clustering loss of a $K$-means clustering algorithm is given by the sum of squares of the distance from each point to the centroids of its cluster, i.e., 
    \begin{equation}\label{eq:clus_loss}
        r(X,\bm{q}) := \frac{1}{P}\sum_{p \in [P]} \|\bm{x}_p - \bm{c}_{\ell_p}\|_2^2, \quad\text{where}\quad \bm{q} = \{\bm{c},\bm{l}\}
    \end{equation}
\end{defn}
We note that this is equivalent to the minimizer of the clustering query in \cref{eq:clus_obj}, and that a lower clustering loss is better. In later sections, we introduce the DP mechanism that will corrupt the query outcome via an additive noise mechanism, and publish $\tilde{\bm{q}}$ in place of $\bm{q}$. This naturally induces an additional loss on top of the clustering loss. Thus, we define the \textbf{\textit{DP (query) accuracy loss}} as follows:
\begin{defn}[DP (query) Accuracy Loss]
    Given a query answer $\bm{q}$ and a DP answer $\tilde{\bm{q}}$, the DP query accuracy loss, or simply the DP accuracy loss, is given by:
    \begin{equation}\label{eq:rel_DP_loss}
        \bar{r}(X,\bm{q},\tilde{\bm{q}}) = \frac{r(X,\tilde{\bm{q}}) - r(X,\bm{q})}{r(X,\bm{q})},
    \end{equation}
    where $r(X,\tilde{\bm{q}})$ is the clustering loss if the outcome of the $K$-means clustering algorithms were $\tilde{\bm{q}}$. Since $\bm{q}$ is the minimizer of \cref{eq:clus}, $\bar{r} \geq 0$, and the equality is met in the absence of a DP mechanism. 
\end{defn}
It is also important to define the \textit{reward} of our DP query answer. The addition of noise strengthens the privacy guarantees while increasing the DP accuracy loss. 
Since the reward and DP accuracy loss have a negative correlation, we will discuss the performance of our mechanisms in terms of their DP accuracy loss alone, with the understanding that lower the DP accuracy loss, the higher is the reward.

Finally, to restate, the goal of our mechanism is to guarantee differential privacy of the $K$-means clustering query answer while minimizing the DP accuracy loss.

\subsubsection{Threat model to the $K$-Means Clustering Query}\label{sec:threat_model}
As discussed in the prior sections, traditional rules of thumb adapted by specific industries have flawed quantification of privacy guarantees, and anonymization often fails in the presence of substantial side information. In this regard, the threat we consider in this paper is that of a third-party analyst's ability to discern the value of any particular data point in the dataset that is under investigation. 

To illustrate the threat, consider a scenario when both the centroids and labels are readily available for a dataset $X$. Now, suppose an additional point $x$ is added to the dataset and this leads to a change in a single centroid, say of a cluster $k$. 
Notice that each centroid is similar to the average query that was under investigation in \Cref{sec:1515}.
Now, using the information about the population of each cluster obtained from the vector of labels, an adversary can infer the value of the point $x$ as described in \Cref{sec:1515}. 

This threat is especially relevant in smart grid data analysis.
To elaborate, consider the average query and a dataset that contains power loads of $P-1$ houses without solar photovoltaics (PVs) and one house with solar PVs installed. During the daytime, there will be periods during which the latter does not consume any power from the electric grid, and in fact injects power to the grid. Then, its load measurement will be negative (by convention) as opposed to the positive load measurements of all the other houses in the dataset. Thus, the average query with and without the last house will be vastly different from one another, making that house highly sensitive to the average query and a possible source of privacy leakage, as now the fact that it contains a solar PV can be inferred by the analyst.

\begin{remark}[Internal and External Threats]
We do not consider insiders (of the organization that stores the data) with legitimately acquired access to the data as threats. 
Instead, we are concerned with the inference of a data point's involvement after a particular aggregate query has been published to an external untrustworthy third party.
\end{remark}

\subsection{Differential Privacy}\label{sec:dp_prelims}
We now introduce DP as an alternative to the widely used ``15/15'' Rule in the smart grid industry to publish private aggregate data. We denote the DP query answer by $\tilde{\mathbb{q}}(X)$, and has a random outcome $\qtilde \in \mathcal{Q}$, with distribution $f(\qtilde|X)$ (the probability density function for continuous random queries and the probability mass function for discrete random variables).
We briefly introduce the conventional definitions that explain how differential privacy is measured and established. The first and the most widespread definition of differential privacy was introduced in \cite{dwork2006calibrating,dwork2006our}. It states that:
\begin{defn}[$(\epsilon,\delta)$-Differential privacy]
A randomized mechanism $\tilde{\mathbb{q}}$ is $(\epsilon,\delta)$-differentially private if for all neighboring datasets $X$ and $X'$ that differ in one point, for any arbitrary event pertaining to the outcome of the query, the randomized mechanism satisfies the following inequality
 \begin{equation}\label{eq:e-dp}
    \forall \mathcal{S},\quad 
          Pr(\tilde{\mathbb{q}}(X) \in \mathcal{S}) \leq \exp(\epsilon)Pr(\tilde{\mathbb{q}}(X') \in \mathcal{S}) + \delta,
\end{equation}
where $Pr(\mathcal{A})$ denotes the probability of the event $\mathcal{A}$, for some privacy budget $\epsilon\geq 0$ and $\delta \in [0,1]$. 
\end{defn}
Note that, since $\delta$ is a bound that may not be tight, smaller values of $\delta$ are possible. Hence, $(\epsilon,\delta)$ guarantees are sufficient but not necessary conditions to ensure that information about $X$ leaks. The authors in \cite{machanavajjhala2008privacy} introduced a revised definition of privacy as follows:
\begin{defn}[$(\epsilon,\delta)$-Probabilistic Differential privacy]\label{def:probabilisticDP} The so-called privacy leakage function  $L_{\XX}$ is the log-likelihood ratio between the two hypotheses that the query outcome $\qtilde$ is the answer generated by the data $X$ or the data $X'$ that differ by one element. Mathematically:
\begin{equation}
    L_{\XX}(\qtilde):=\log \frac{f(\qtilde|X)}{f(\qtilde|X')},
\end{equation}
  A randomized mechanism $\tilde{\mathbb{q}}(X)$ is $(\epsilon,\delta)$ differentially private for $X$ if and only if:
  \begin{equation}\label{eq:def1}
      \sup_{X'}~Pr\left(L_{\XX}(\qtilde) >\epsilon\right)\leq \delta.
  \end{equation}
\end{defn}
It can be shown that $(\epsilon,\delta)$-PDP is a strictly stronger condition than $(\epsilon,\delta)$-DP. 
\begin{thm}[PDP implies DP~\cite{mcclure2015relaxations}]\label[thm]{thm:PDP-DP}
If a randomized mechanism is $(\epsilon,\delta)$-PDP, then it is also $(\epsilon,\delta)$-DP, i.e.,
\[
    (\epsilon,\delta)-\text{PDP}  \Rightarrow (\epsilon,\delta)-\text{DP}, \text{ but } (\epsilon,\delta)-\text{DP} \nRightarrow (\epsilon,\delta)-\text{PDP}.
\]
\end{thm}

In dealing with a multidimensional answer $\qtilde$ (as is for the case of a $k$-means clustering algorithm that returns $K$ centroids and $P$ data labels) a first common simplification is to use independent mechanisms for different components; a second common simplification is to use the following lemma to map the $(\epsilon_j,\delta_j)$ results for the scalar independent mechanism employed onto a global $(\epsilon,\delta)$ result, rather than  $L_{\XX}(\qtilde)=\sum_{j=1}^m L_{\XX}(\tilde{q}_j)$.
\begin{lemma}[Sequential composition~\cite{dwork2014algorithmic}]\label[lemma]{lem:series_composition}
If $n$ randomized algorithms $M_i$, $i=1,2,\ldots,n$, are $(\epsilon_i,\delta_i)$-DP, then the sequential execution of these algorithms on the database $X$ provides $(\sum_i \epsilon_i, \sum_i \delta_i$)-differential privacy.
\end{lemma}

In this paper, we use \Cref{def:probabilisticDP} which has an immediate statistical interpretation: if values of $(\epsilon,\delta)$ are close to zero, even when one adopts the optimum statistical test for the hypotheses that the randomized answer is produced by the neighboring datasets $X$ or $X'$ that differ in one point, the test produces results that mostly are incorrect or are unreliable answers (i.e., there is a non-zero probability that the test is wrong). Of course, this comes at a cost in terms of accuracy of the answer, which is important to account for. 

Next, we describe the mechanism $\tilde{\mathbb{q}}(X)=\{\tilde{\mathbb{c}}(X),\tilde{\mathbb{l}}(X)\}$ that renders the clustering query answer $(\epsilon,\delta)$-DP.


\section{An \texorpdfstring{$(\epsilon,\delta)$}{(e,d)}-DP Mechanism for \texorpdfstring{$K$}{K}-Means Clustering }\label{sec:additive}
The $K$-means clustering query, as described in \ref{sec:kmeans_prelims}, has two components: the centroids and the labels ${\mathbb{q}}(X)~=~\{{\mathbb{c}}(X),{\mathbb{l}}(X)\}$. Our method introduces independent randomized mechanisms applied to each component and uses composition as defined in \Cref{lem:series_composition} to provide overall DP guarantees.

\subsection{Differential Privacy Guarantees for Cluster Centroids}\label{sec:centroids_additive}
In this section, we first remind the readers of the classical white Gaussian noise mechanism and then introduce the novel additive colored Gaussian noise mechanism, where we perturb the output by adding noise to the cluster centroids found by an appropriate clustering algorithm with convergence guarantees. 

As discussed in \Cref{sec:1515}, any DP mechanism should introduce uncertainty on the query answer so that its random outcomes under two neighboring datasets $X$ or $X'$ are statistically indistinguishable with high probability.

\subsubsection{Review of the White Gaussian Noise Mechanism}\label{sec:white_gaussian}
\begin{figure}[!htbp]
    \centering
    \includegraphics[width=0.45\textwidth]{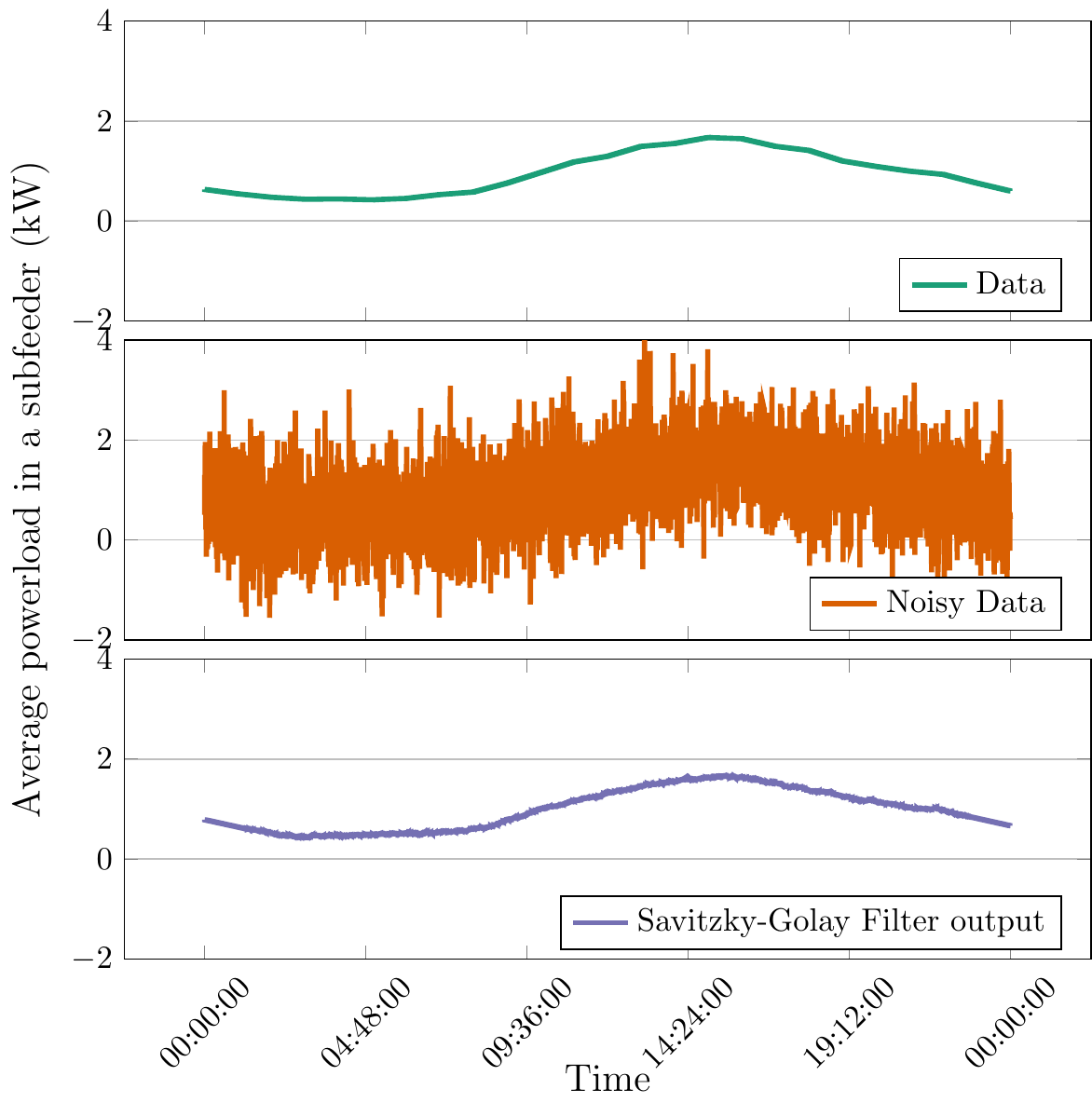}
    \caption{Illustration of the pitfalls of white noise mechanisms on smooth time-series data.}
    \label{fig:filterting}
\end{figure}

The Gaussian noise output perturbation mechanism is a popular option in DP for publishing a variety of statistics. The approach entails adding i.i.d. noise to the cluster centroids as follows:
\begin{equation}
    \tilde{\mathbb{c}}(X) = \mathbb{c}(X) + \bm{\eta} = \bm{c} + \bm{\eta},\label{eq:dp_centroid}
\end{equation}
where $\bm{\eta}$ is a normally distributed noise vector with mean $\bm{0}$ and variance $\sigma^2\bm{I}$. The following theorem provides the DP guarantees for such a method:
\begin{thm}[Cluster centroids are $(\epsilon_c,\delta_c)$-DP~\cite{dwork2006calibrating}]\label[thm]{thm:cluster_centroids_DP}
The additive noise mechanism in \cref{eq:dp_centroid} provides $(\epsilon_c,\delta_c)$-DP for any two neighboring datasets $X$ and $X'$ when $\bm{\eta}~\sim~\mathcal{N}(\bm{0},\sigma^2\bm{I})$, for some $\epsilon_c \geq 0$ and $\delta_c \in [0,1]$, where 
\begin{equation}
    \sigma \geq \frac{\Delta\mathbb{c}}{\epsilon_c} \sqrt{2\log(2/\delta_c)}\label{eq:color_gaussian}
\end{equation}
and $\Delta\mathbb{c}$ is the query sensitivity given by:
\begin{equation*}
\Delta\mathbb{c} = \sup_{X'}\|\mathbb{c}(X) - \mathbb{c}(X')\|_2.    
\end{equation*}
\end{thm} 
But it can be shown that a white noise mechanism applied to a smooth time-series dataset can be filtered out. For example, in \cref{fig:filterting}, we plot the average of the daily power loads of the houses in a feeder along with a DP white Gaussian noise treated load curve, where noise standard deviation was set at $\sigma = 0.75$. This noise can easily be filtered out by a Saviztky-Golay filter \cite{press1990savitzky} (of order $1$ with a window size of $300$), a generalized moving average mechanism, as shown in bottom plots in \cref{fig:filterting}.
For a dataset containing the power loads of houses in a set of feeders, the centroids found through a $K$-means algorithm are often in the low-frequency domain, and adding white noise that reside in the high-frequency domain is often not enough.

\subsubsection{Proposed Colored Gaussian Noise Mechanism}
As an alternative to the conventional method of adding i.i.d. noise to each entry of the query, we consider the fact that, particularly in the case of time series, it is often the case that they are sparse in a transform domain, such as the Discrete Fourier Transform or the Wavelet transform. In such a scenario, we propose the addition of colored Gaussian noise instead, where the noise added depends on the relative positions of the centroids in question:
\begin{equation}
    \tilde{\mathbb{c}}(X) = \mathbb{c}(X) + \hat{\bm{\eta}} ~~\mbox{where}~~\hat{\bm{\eta}} \sim \mathcal{N}(\bm{0}, \bm{\Gamma}^{-1}),\label{eq:dp_centroid_color}
\end{equation}
where 
where $\bm{\Gamma}$ is the noise precision matrix. 
The following theorem states the privacy guarantees of this mechanism, with its proof presented in~\cite[Theorem 4]{ravi2021colored}.
\begin{thm}[Colored Gaussian Noise Mechanism is $(\epsilon_c,\delta_c)$-DP]\label[thm]{thm:cluster_centroids_DP_color}
The additive noise mechanism in \cref{eq:dp_centroid_color} provides $(\epsilon_c,\delta_c)$-DP for any two neighboring datasets $X$ and $X'$ that differ in one point, for some $\epsilon_c \geq 0$ and $\delta_c \in [0,1]$.
\end{thm} 
While the proof is presented in \cite{ravi2021colored}, we briefly provide the sketch of the proof here.
\begin{proof}
    The design of the optimal noise vector hinges on the design of its covariance matrix. This is seen by analyzing the numerator of the DP accuracy loss function in \cref{eq:rel_DP_loss}, which quantifies the deviation of the accuracy loss function (defined in \cref{eq:clus_loss}) at the noisy query answer $\tilde{\bm{q}}$ from its optimal value obtained at the true query answer $\bm{q}$. Under the Gaussian mechanism, the clustering loss is given by:
\begin{align}
    r&(X,\tilde{\bm{q}}) = \frac{1}{P}\!\!\!\sum_{p \in [P]}\!\!\|\bm{x}_p - \tilde{\bm{c}}_{\ell_p}\|_2^2 = \frac{1}{P}\!\!\!\sum_{p \in [P]} \|\hat{\bm{\eta}}_{\ell_p}- \left(\bm{x}_p - \bm{c}_{\ell_p}\right)\|_2^2\nonumber\\
    &= r(X,\bm{q}) + \sum_{k\in[K]}\|\hat{\bm{\eta}}_k\|_2^2 - \frac{2}{P}\sum_{p \in [P]} \langle \hat{\bm{\eta}}_{\ell_p}, \bm{x}_p - \bm{c}_{\ell_p}\rangle\nonumber\\
    &= r(X,\bm{q}) + \|\hat{\bm{\eta}}\|_2^2 - \frac{2}{P}\sum_{p \in [P]} \langle \hat{\bm{\eta}}_{\ell_p}, \bm{x}_p - \bm{c}_{\ell_p}\rangle,\nonumber
\end{align}
where $\hat{\bm{\eta}} = [\hat{\bm{\eta}}_0^T,\ldots,\hat{\bm{\eta}}_{K-1}^T]^T$ with $\hat{\eta}_k$ being the noise vector added to the centroid of cluster $k$. Thus, we can write the following:
\begin{equation*}
    \mathbb{E}_{\hat{\bm{\eta}}}[r(X,\tilde{\bm{q}}) - r(X,\bm{q})] = \mathrm{Tr}(\mathrm{covar}(\hat{\bm{\eta}})) = \mathrm{Tr}(\bm{\Gamma}^{-1}),
\end{equation*}
where $\mathrm{Tr}(\bm{A})$ is the trace of the matrix $\bm{A}$ given by the sum of the elements on its diagonal. Hence, in order to achieve the lowest DP accuracy loss, we have to minimize the trace ($\mathrm{Tr}$) of the noise covariance matrix (the inverse of $\bm \Gamma$).

Concurrently, in order to maintain the DP guarantees, we need to satisfy \cref{eq:color_gaussian}. Thus, we have the following condition:
\begin{equation}
    \Delta\mathbb{c}^2 \leq \frac{\epsilon_c^2}{2\log(2/\delta_c)} =: \gamma_c,
\end{equation}
where $\sigma$ is set to $1$, and the cluster query sensitivity is:
\begin{equation}
    \Delta\mathbb{c} = \sup_{X'} \|\bm{\Gamma}^{1/2}\left(\mathbb{c}(X)\!-\!\mathbb{c}(X')\right)\|_2.
\end{equation}
Thus, for all neighboring datasets $X' \in \mathcal{X}$ of $X$ that differ in one element, we have that:
\begin{align*}
    \left(\mathbb{c}(X) - \mathbb{c}(X')\right)^{\T} \bm{\Gamma}\left(\mathbb{c}(X) - \mathbb{c}(X')\right) \leq \Delta\mathbb{c}^2 \leq \gamma_c.
\end{align*}

Finally, we write the following optimization problem to find the optimal noise covariance matrix that finds the covariance matrix that satisfies the DP conditions while providing the least DP accuracy loss:
\begin{align}
    \min_{\bm{\Gamma}}~&\tr{\bm{\Gamma}^{-1}}\label{eq:color_optimization}\\
    \text{s.t.}~&\left(\mathbb{c}(X)-\mathbb{c}(X')\right)^{\T} \bm{\Gamma} \left(\mathbb{c}(X)-\mathbb{c}(X')\right) \leq \gamma_c, \hfill \forall X'\in {\cal X}.\nonumber
\end{align}
In the following lemma, we provide the means to find the optimal noise covariance (see ~\cite[Lemma 5]{ravi2021colored} for proof):
\begin{lemma}[Optimal Choice of $\bm{\Gamma}$]\label[lemma]{lem:optimal_Phi}
Let the matrix $\bm C_{\XX}$ contain as its columns all possible $(\mathbb{c}(X)-\mathbb{c}(X'))$, $X'\in {\cal X}$ and let us assume that $\bm C_{\XX}$ is full row rank, and that the first $Kd$ columns of $\bm C_{\XX}$, corresponding to the set ${\cal D}\subseteq \cal X$ have the smallest norms and are linearly independent, forming the matrix we refer to as $\bm C_{\XX}^*$. Then, the optimization problem in \cref{eq:color_optimization} has a unique solution  and it evaluates to:
\begin{equation}
    \bm{\Gamma}^\star = {\bm{R}}_{\bm{\lambda}^\star}^{-\frac{1}{2}},\label{eq:opt_col_noise_cov}
\end{equation}
where $\bm{\lambda}^*$ is the vector consisting of the Lagrangian multipliers associated with the optimization problem and has only $Kd$ non-zero values which correspond to the constraints associated with the set ${\cal D}$, and:
\begin{equation}
    \bm{R}_{\bm{\lambda}^\star}\!:=\!\sum_{X' \in \mathcal{D}} {\lambda}_{X'}^\star \left(\mathbb{c}(X)\!-\!\mathbb{c}(X')\right)\left(\mathbb{c}(X)\!-\!\mathbb{c}(X')\right)^{\T},
\end{equation}
where $\lambda_{X'}^\star, \forall X'\in {\cal D}$ are the non-zero Lagrange multipliers for the problem in \cref{eq:color_optimization}. Their values are:
\begin{align}
    \lambda^*_i =v_i^{-2} ~~\mbox{where}~~\bm v=\gamma_c\bm M^{-1}\bm 1, ~~ M_{ij}=[\bm C_{\XX}^{*\frac{\T}{2}}]^2_{ij}.
\end{align}
\end{lemma}

Thus, the mechanism in \cref{eq:dp_centroid_color} is ($\epsilon_c,\delta_c$)-DP with the noise covariance set according to \cref{eq:opt_col_noise_cov}.
\end{proof}

\subsection{Differential Privacy Guarantees for the Labels}\label{sec:labels_additive}
In this section, we focus on a mechanism for publishing the labels in a way that is differentially private.
This task is less straightforward than the previous one. To begin with, if the number of points $P$ is extremely high compared to the number of clusters $K$, then the removal of one point from the dataset might not necessarily change the labels of the rest of the points, it is unnecessary to randomize the labels. We have to ensure that we randomize the labels of a subset of the points, which when removed changes the labels of the rest of the points, i.e., those point which are sensitive to the label query. Secondly, to complicate matters, adding random errors to the labels of points very close to the centroids would result in extremely inaccurate results, making the answers completely useless. 
Given these two caveats, the task of any algorithm is two-fold:
\begin{enumerate}[leftmargin=*]
    \item[i)] Choose the ideal subset of points whose labels are to be randomized, and 
    \item[ii)] randomize the labels of these points such that we achieve the least error.
\end{enumerate}

\subsubsection{Choice of Points to Randomize} \label{sec:label_choice}
In a densely populated field of points, the effect of the removal of one point from the dataset on the position of the centroids is minimal. Thus, points closer to their centroids and points in clusters far away from other clusters tend to retain their labels. However, the labels of the points in the periphery of a cluster, and especially those in the vicinity of points from another cluster, might flip on the removal of a point from the dataset. Thus, only these edge points contribute to the sensitivity of the label query. The mechanism that we propose in this section only adds label noise to a subset of the population, say $\mathcal{L} \subseteq [P]$, which is the set of all the points whose label changes if any other point is removed. The composition of $\mathcal{L}$ is illustrated in \cref{fig:mathcal{L}}, where we have a set of points $\{0,\ldots,19\}$. Here, when point $1$ is removed from the set, the new cluster affiliations of the points in $\{2,7,17\}$ are different from what they were before. Similarly, the cluster affiliations of points in $\{2,9,13\}$ (respectively $\{4,15\}$) are changed when point $5$ (respectively point $14$) is removed. The union of all such sets of points forms the set $\mathcal{L}$.
\begin{figure}
    \centering
    \includegraphics[width=0.45\textwidth]{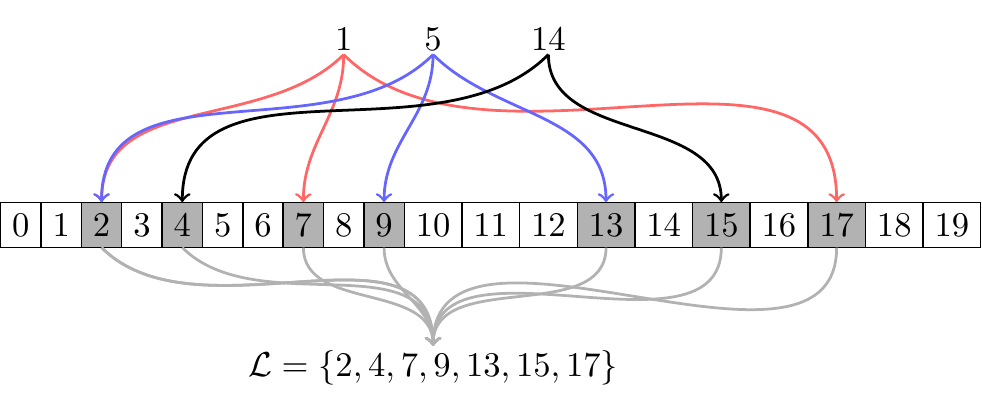}
    \caption{The composition of the set of points whose labels are treated with the DP Label mechanism.}
    \label{fig:mathcal{L}}
\end{figure}

\subsubsection{Randomized Mechanism for the Labels}
Unlike in the case of cluster centroids, the use of Gaussian (or any other unbounded noise) mechanism is impermissible for the labels, as they are integers belonging to the set $[K]$. Thus, we employ a modulo-$K$ additive mechanism where the output is given by:
\begin{equation}
    [\tilde{\mathbb{l}}(X)]_p = \begin{cases}
        [\mathbb{l}(X)]_p \oplus_K \nu_p, & p\in \mathcal{L}\\
        [\mathbb{l}(X)]_p, & \text{otherwise},
        \end{cases}\label{eq:dp_label}
\end{equation}
where $\{\nu_p\}_{p\in\mathcal{L}}$ are i.i.d random variables in the set $[K]$ and $\oplus_K$ is the modulo $K$ addition operator, which ensures that the range remains in the permissible set $[K]$. The mechanism involves the modular addition of i.i.d noise samples to the labels of each of the points in $\mathcal{L}$. In the following theorem, we provide the DP guarantees afforded by this mechanism with the proof in \Cref{app:label_DP}.

\begin{thm}[Labels are $(\epsilon_\ell,\delta_\ell)$-DP]\label[thm]{thm:label_DP}
The modulo additive noise mechanism in \cref{eq:dp_label} provides $(\epsilon_\ell,\delta_\ell)$-DP for any two neighboring datasets $X$ and $X'$ when $\{\nu_p\}_{p\in\mathcal{L}}$ are i.i.d and have the following probability mass function:
\begin{equation*}
    f(\nu_p) = \begin{cases}
        1-\rho, & \nu_p = 0\\
        \frac{\rho}{K-1}, & \nu_p = [K]\setminus \{0\},
    \end{cases}
\end{equation*}
where  $K \geq 2$, $\rho < 0.5$, $\Delta\mathbb{l}$ is the query sensitivity given by:
\begin{equation*}
    \Delta\mathbb{l} = \sup_{X'} \|\mathbb{l}_{\mathcal{L}}(X') - \mathbb{l}_{\mathcal{L}}(X)\|_0,
\end{equation*}
and $\delta_\ell$ is $0$ when $\epsilon_\ell > \Delta\mathbb{l} \times \log\left(\frac{(1-\rho)(K-1)}{\rho}\right)$ and when $\epsilon_\ell < \Delta\mathbb{l} \times \log\left(\frac{(1-\rho)(K-1)}{\rho}\right)$, it is given by
\begin{align*}
    \delta_\ell &= \Delta\mathbb{l}! \left[ \frac{\rho(K-2)}{K-1}\right]^{\Delta\mathbb{l}} \times \nonumber\\
    & \sum_{m_c = 0}^{\Delta\mathbb{l}}\sum_{m_0 = \ell + m_c}^{\Delta\mathbb{l}} \frac{\left[ \frac{(K-1)(1-\rho)}{\rho} \right]^{m_0} (K-2) ^{-(m_0+m_c)}}{m_0! m_c! (\Delta\mathbb{l} - m_0 - m_c)!}.
\end{align*}
\end{thm}

We measure the label mechanism's degradation of the solution as the number of points whose labels are changed, i.e., $\|\mathbb{l}_{\mathcal{L}}(X) - \tilde{\mathbb{l}}_{\mathcal{L}}(X)\|_0$, and in expectation this loss is given by:
\begin{equation*}
    \mathbb{E}[\|\mathbb{l}_{\mathcal{L}}(X) - \tilde{\mathbb{l}}_{\mathcal{L}}(X)\|_0] = \mathbb{E}_{\bm{\nu}}\left[\textstyle{\sum_{p \in \mathcal{L}}} \mathbbm{1}_{\{\nu_p \ne 0 \}}\right] = |\mathcal{L}| \rho.
\end{equation*}
Finally, the total degradation of the combined centroid and label perturbations is measured as follows:
\begin{equation*}
    r(X,\tilde{\bm{q}}) := \frac{1}{P}\sum_{p\in[P]} \|\bm{x}_p - \tilde{\bm{c}}_{\tilde{\ell}_p}\|_2^2.
\end{equation*}

\subsection{Mechanism and its Differential Privacy Guarantees}\label{sec:overall_additive}
In \Crefrange{sec:centroids_additive}{sec:labels_additive}, we provided differential private mechanisms for the publication of cluster centroids and point labels. In \cref{alg:col_lab_mech}, we provide the overall DP $K$-means mechanism with colored Gaussian noise for the centroids and discrete noise for the labels. 
\begin{algorithm}[!htbp]
    \caption{DP Mechanism with colored Gaussian noise and randomized labels}\label{alg:col_lab_mech}
    \begin{algorithmic}[1]
    \Procedure{DpK-means}{$\mathbbm{q},X,\epsilon_c,\delta_c,\epsilon_{\ell},\delta_{\ell}$}
    \State $\bm{q}(\bm{c},\bm{\ell}) \gets$ Cluster $X$ into $K$ clusters.
    \State Calculate $\bm{\Gamma}$ according to \cref{eq:opt_col_noise_cov} with $\epsilon_c$ and $\delta_c$.
    \State $\tilde{\bm{c}} \gets \bm{c} + \bm{\eta}$, where $\bm{\eta} \sim \mathcal{N}(\bm{0},\bm{\Gamma})$.
    \State $\tilde{\bm{\ell}} \gets \bm{\ell} \oplus \bm{\nu},$ with $\bm{\nu}$ distributed according to the conditions of \cref{thm:label_DP}.
    \EndProcedure
    \end{algorithmic}
    \hspace*{\algorithmicindent} \textbf{Output:} $\tilde{\bm{q}} = \{\tilde{\bm{c}},\tilde{\bm{\ell}}\}$. 
    \end{algorithm}
In the following corollary, we state the overall DP guarantees that encapsulates the complete query as described in \cref{eq:clustering_components}.
\begin{cor}[Clustering query is $(\epsilon,\delta)$-DP]\label[cor]{thm:overall_DP}
For any two neighboring datasets $X$ and $X'$, the mechanisms described in \cref{eq:dp_centroid} and \cref{eq:dp_label} together provide $(\epsilon,\delta)$-DP guarantees with:
\begin{equation}
    \epsilon = \epsilon_c + \epsilon_\ell, \quad \delta = \delta_c + \delta_\ell.
\end{equation}
\end{cor}
The proof follows from \Crefrange{thm:cluster_centroids_DP_color}{thm:label_DP} and \Cref{lem:series_composition}.

\section{Use Case: Power Systems Aggregate Query Publication and Synthetic Load Generation}\label{sec:use-case}

The guarantees provided by the proposed DP clustering mechanism are tailor-made to the ones that regulatory agencies are trying to achieve with regard to the publication of consumer data,
since third-party agencies will not be able to gleam information about the presence, or lack thereof, of any individual entry in the database. 

The method proposed is general and can be applied to any kind of data. In this section, 
we go one step further and use the guarantees afforded in the publication of labels and centroids to generate synthetic load profiles for the nodes in the dataset, leveraging an empirical property that is specific to AMI data. In fact, it so happens that the daily load profiles when divided into clusters (to be discussed in \cref{sec:real_data} in detail) exhibit an excellent fit with a multivariate log-normal distribution whose generation is relatively simple. The first step to generate synthetic samples is to group the dataset by clusters, take the logarithm (adding a large constant if the net load data have negative values) and then use the DP centroids and DP covariance to fit separate Gaussian distributions to each cluster and then go back to power loads profiles taking an exponential (and subtracting the constant, if needed. 
\begin{algorithm}
\caption{Algorithm to generate Load Profiles}\label{alg:lognorm}
\begin{algorithmic}[1]
\Procedure{logNormSamples}{}
\State Cluster $X$ into $K$ clusters.
\For{$k \in [K]$}
\State $\bm{X}_k \gets {\log}([\bm{x}_{p_1} \ldots \bm{x}_{p_{|\tilde{\mathcal{C}}_k|}}] + \alpha \bm{1}\bm{1}^{\T})$
\State $\hat{\bm{\mu}}_k = \frac{1}{|\tilde{\mathcal{C}}_k|} \bm{X}_k\bm{1} +  \bm{\eta}_k$ 
\State $\hat{\bm{\Sigma}}_k = \frac{1}{|\tilde{\mathcal{C}}_k|} \bm{X}_k\bm{X}_k^{\T} + \bm{\Phi}_k$
\State $\hat{\bm{X}} \gets $ \text{samples drawn from } $\mathcal{N}(\hat{\bm{\mu}}_k,\hat{\bm{\Sigma}}_k)$
\State $\hat{\bm{X}} \gets {\exp}(\hat{\bm{X}}) - \alpha \bm{1}\bm{1}^{\T}$.
\EndFor
\EndProcedure
\end{algorithmic}
\end{algorithm}
Letting $\exp(X)$ (resp. $\log(\bm{X})$) denote the matrix where each element is the exponent (resp. logarithm) of the corresponding element in $\bm{X}$, the algorithm in \cref{alg:lognorm} generates prototypical load shapes for each cluster. Our DP clustering mechanism renders the estimates of the means and labels differentially private. To obtain a DP covariance, we apply the Wishart mechanism~\cite{jiang2016wishart}. Using this fitted distribution, we then generate multivariate Gaussian samples, before finally outputting the exponential of the generated samples. This approach is summarized in \cref{alg:lognorm}, where the $\alpha$ parameter is used to ensure that the argument of the log operation remains positive, and $\bm{\eta}_k$ and $\bm{\Phi}_k$  are the noise added to the cluster centroid and to the covariance matrix of the data pertaining to the cluster $k$. 

\section{Numerical Results}\label{sec:numericals}
In this section, the proposed methods are tested on a synthetic dataset consisting of points drawn from a Gaussian mixture and then with two real-world datasets.

\subsection{Synthetic Dataset}

\begin{figure}
    \centering
    \includegraphics[width=0.4\textwidth]{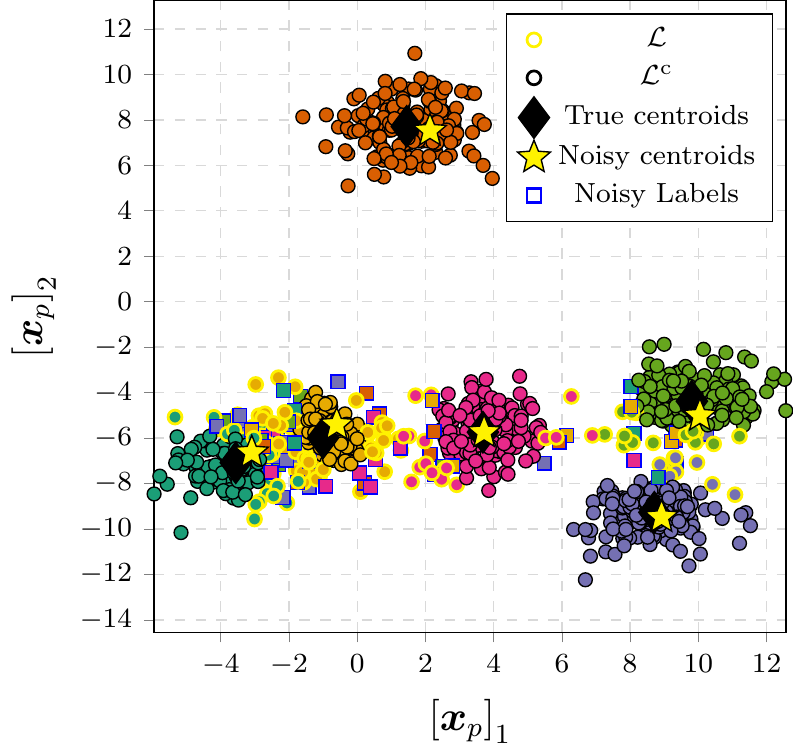}
    \caption{Gaussian Mixture Dataset scatter plots: the six classes are indicated using different colors, with the points belonging to the set $\mathcal{L}$ indicated by smaller marks with a yellow border. Of these, the points whose labels changed are indicated using square ({\color{blue}$\Box$}) markers.}
    \label{fig:g_mix_scatter}
\end{figure}
\begin{figure*}
    \centering
    \includegraphics[width=\textwidth]{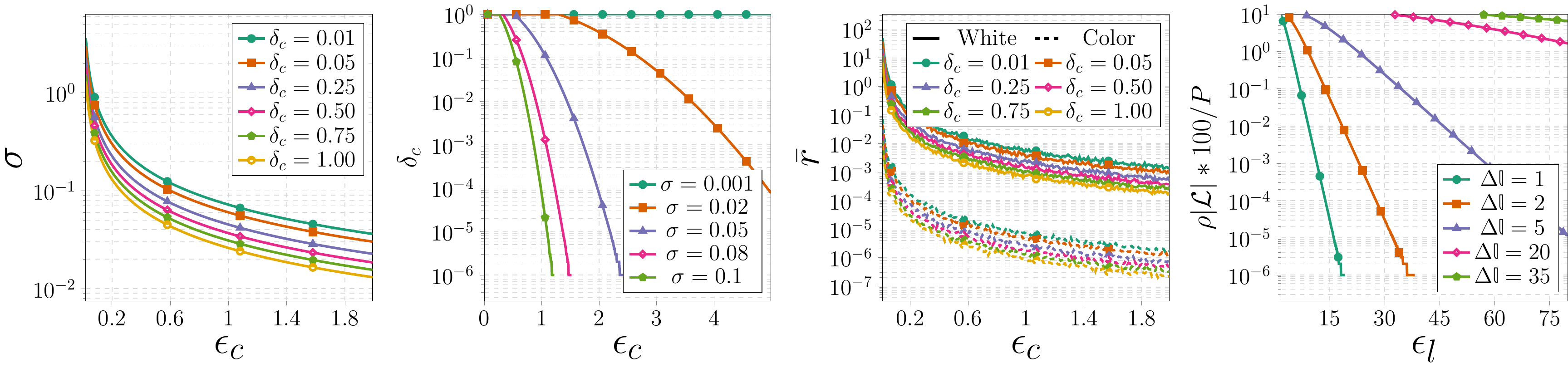}
    \caption{DP performance for Gaussian Mixtures.}
    \label{fig:g_mix_perf}
\end{figure*}
In this section $X$ consists of $P = 1000$ points, each randomly drawn from a  Gaussian mixture with $K = 6$ equally likely components that are in $\mathbb{R}^2$ and are ${\cal N}(\bm \mu_k,\sigma\bm I)$ $k\in [K]$ with $\bm \mu_k\in\mathbb{R}^2$. Note that because the mixture components covariance is $\bm I$, here we do not need colored noise for the centroids.  The output of the $K$-means clustering algorithm is shown in \cref{fig:g_mix_scatter} and it is the clustering assignment that we consider as the true assignment. In the same figure, we highlight the elements of the set $\mathcal{L}$ by marking them with markers encircled by yellow borders. These points have additive noise added to their labels, as shown in \cref{eq:dp_label}, and those whose labels changed after the DP mechanism are highlighted with square ($\Box$) markers with blue borders. In \cref{fig:g_mix_perf}, we show the variation of noise variances for various $\delta_c$s, the variation of $\delta_c$ for various noise variances and the DP accuracy loss for various $\delta_c$s. As expected, in order to achieve a stronger privacy guarantee (i.e., lower $\delta_c$ values for given $\sigma$), we require a higher privacy budget and, in a similar vein, a lower DP accuracy loss (in other words, a lower noise variance) requires a higher privacy budget for a given level of privacy guarantee. It is also important to note the effect of the sensitivity on the performance of the DP mechanism proposed for the labels. Queries that are highly sensitive to the underlying dataset require a larger noise variance and, in turn, reduce the reward of the DP query answer.
This is illustrated in \cref{fig:g_mix_perf}~(right), where, as the sensitivity $\Delta \mathbb{l}$ increases, the expected percent of errors, $\rho|\mathcal{L}|*100/P$, also increases for a given privacy budget.

\subsection{Real-World Power Systems AMI Dataset}\label{sec:real_data}
\begin{figure}
    \centering
    \includegraphics[width=0.4\textwidth]{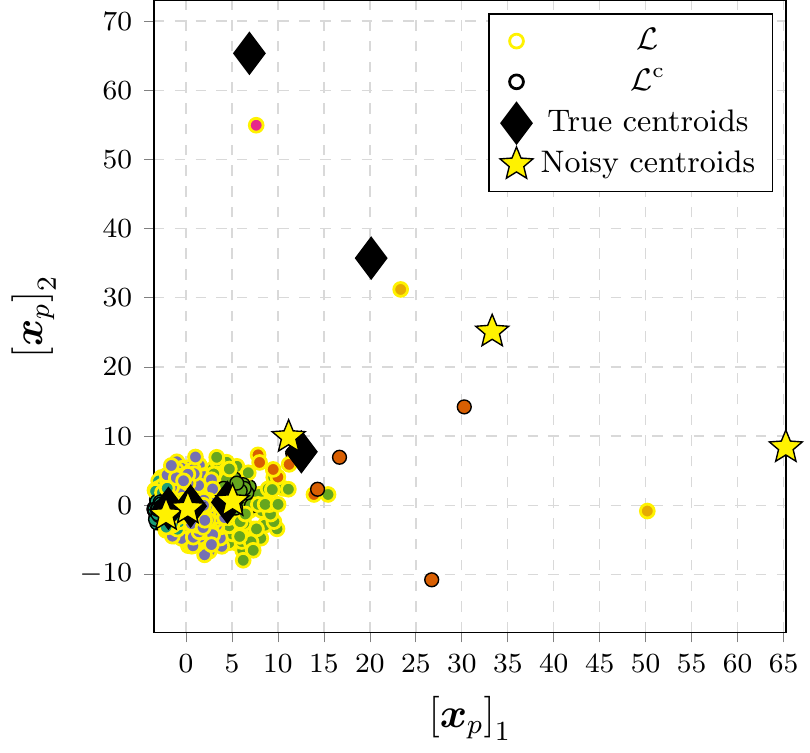}
    \caption{Scatter plots of AMI dataset mapped in 2-$d$ using MDS. The high sensitivity $\Delta\mathbb{c}~=~39.20$ is due to outliers.}
    \label{fig:ami_sum_scatter}
\end{figure}

In this section, we apply our mechanisms for the publication of $K$ means statistics for $P = 1416$ houses' daily profiles (with one hour-resolution) that belong to $12$ distribution circuits across California, USA. The dimension of each house's daily profile, $\bm{x}_p$, is $d=25$ accounting for the load consumed from midnight to midnight. With a choice of $K=6$ clusters, to visualize the daily profiles that are in $\mathbb{R}^{25}$, we map them on $\mathbb{R}^2$ using Multidimensional Scaling\footnote{MDS is a form of non-linear dimensionality reduction which is used to translate information about the pairwise `distances' among a set of $n$ objects or individuals into a configuration of $n$ points mapped into an abstract Cartesian space \cite{mead1992review}. It is important to note that points shown in MDS are solely for visualization purposes and not to be interpreted as containing behind-the-meter generation if a node's co-ordinates are negative.} (MDS) and, as before, in \cref{fig:ami_sum_scatter}, the points in $\mathcal{L}$ are highlighted using markers encircled in yellow borders.
As seen in the MDS plot in \cref{fig:ami_sum_scatter}, we have close to $7$ points that are clear outliers in the dataset.
These outliers drive up the sensitivity value, which would require either higher noise covariance or a very large DP budget.

\begin{figure}
    \centering
    \includegraphics[width=0.4\textwidth]{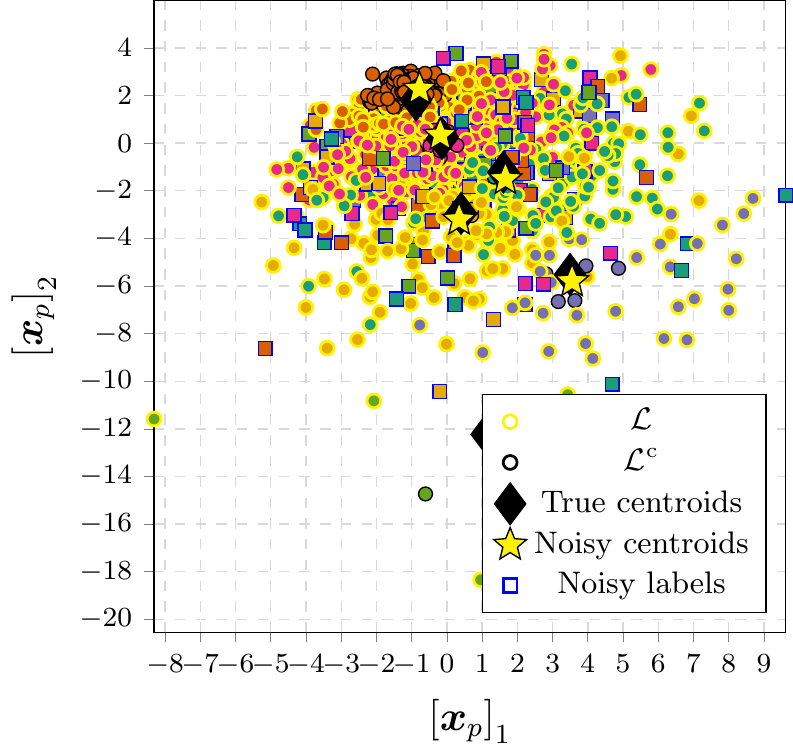}
    \caption{Scatter plot of the AMI Dataset with outliers omitted, which has a sensitivity $\Delta\mathbb{c} = 2.13$. Note that the points, which are originally 25d, are embedded in a 2d-space using MDS. $\epsilon_c = 30, \delta_c = 0.2, \epsilon_{\ell}=30, \delta_{\ell}=0$.}
    \label{fig:ami_sum_filtered_scatter}
\end{figure}
\begin{figure*}
    \centering
    \includegraphics[width=1\textwidth]{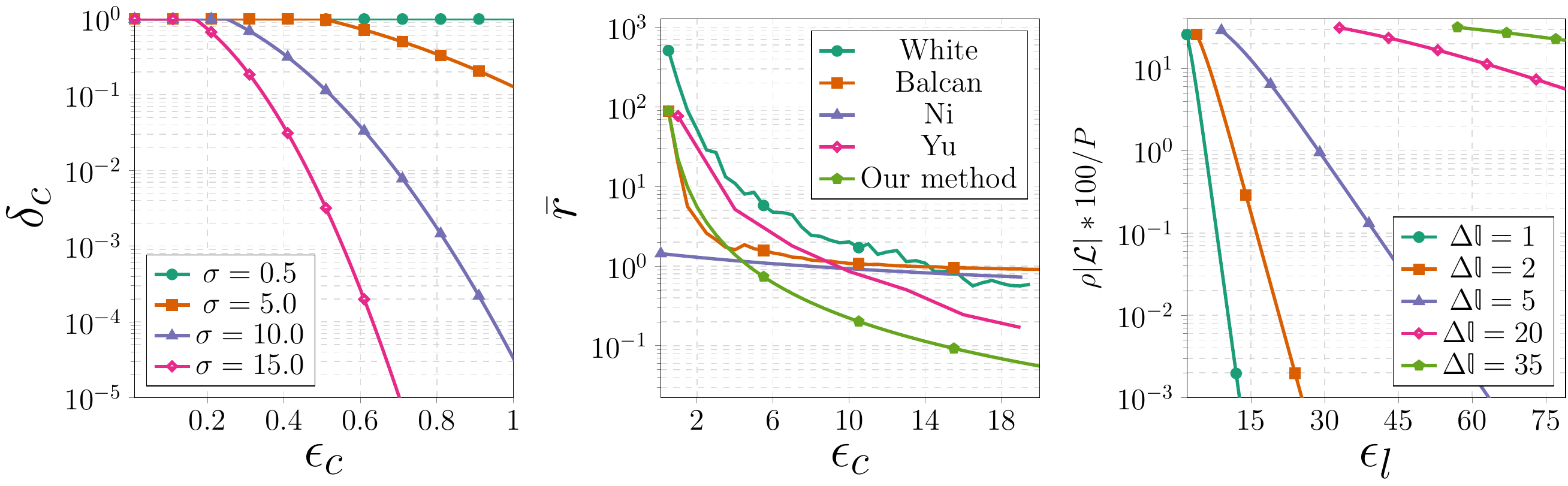}
    \caption{Performance guarantees for the AMI Dataset with outliers omitted, which has a sensitivity of $\Delta\mathbb{c} = 2.13$. (Left) A plot of $\epsilon_c$ vs $\delta_c$, (center) a plot of $\epsilon_c$ vs DP accuracy loss ($\delta_c=0.01$), where our colored Gaussian mechanism with label noise is compared with the methods proposed by Balcan et al. \cite{balcan2017differentially}, Yu et al. \cite{yu2016outlier}, Ni et al. \cite{ni2021utility} and the white Gaussian noise mechanism, and (right) a plot of $\epsilon_\ell$ vs Expected percent of error in labels.}
    \label{fig:ami_sum_filtered_perf}
\end{figure*}

\underline{Domain Specific Knowledge}: 
It is extremely important to pair the guarantees provided by DP with domain specific knowledge in order to extract the best possible reward. With the assumption that the data holder has knowledge regarding the class (i.e., commercial, residential, farm, etc.) to which each node belongs to and depending on the application for which the data is being utilized, it is prudent that we analyze nodes that belong to similar classes and exclude class outliers. For example, in order to incentivize desired behavior by their customers, utilities might compare consumers with similar contracts, connected to distribution circuits in similar areas and climates and not mix commercial and residential customers. The clustering problem of the reduced dataset with outliers (commercial customers on closer inspection) removed, now with $P=1409$, has a sensitivity equal to $2.13$, which has reduced by a factor close to $20$. We show the scatter plot of the dataset (using MDS) in \cref{fig:ami_sum_filtered_scatter} with the true and noisy cluster centroids, where a colored Gaussian noise with $\epsilon_c = 30, \delta_c = 0.2, \epsilon_{\ell}=30, \delta_{\ell}=0$. In addition, the points whose labels were modified are also indicated using a $\square$ marker. We show the $(\epsilon_c,\delta_c)$ tradeoff in \cref{fig:ami_sum_filtered_perf} (left), and in \cref{fig:ami_sum_filtered_perf} (center), we compare our method against the white noise mechanism and the ones discussed by Balcan \cite{balcan2017differentially}, Yu \cite{yu2016outlier}, and Ni \cite{ni2021utility}. It shows a clear improvement in the DP accuracy loss for a given $(\epsilon_c,\delta_c)$ pair for our method compared to the others. The method proposed in \cite{yu2016outlier} first eliminates the outliers as a preprocessing step, similar to our preprocessing step above, but differs in the noise mechanism used to perturb the centroids, and does not provide privacy guarantees on published labels. In fact, the $20$-nearest neighbors based outlier elimination heuristic in \cite{yu2016outlier} removes the same data points that we removed from our original dataset. Note that the DP schemes that utilize a Laplacian noise mechanism all have $\epsilon$-DP guarantees, while our scheme has an $(\epsilon,\delta)$-DP guarantee. In essence, with a small increase $\delta$ above $0$, our mechanism provides a far better DP accuracy loss performance.


\begin{figure*}
    \centering
    \includegraphics[width=1\textwidth]{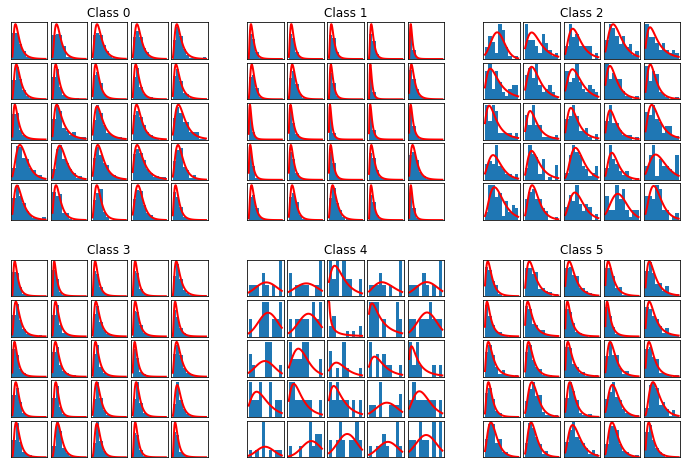}
    \caption{Histograms of the load profiles at each time interval, grouped by cluster. The red curves show the best fit log-normal PDF.}
    \label{fig:hists}
\end{figure*}
\begin{figure*}
    \centering
    \includegraphics[width=1\textwidth]{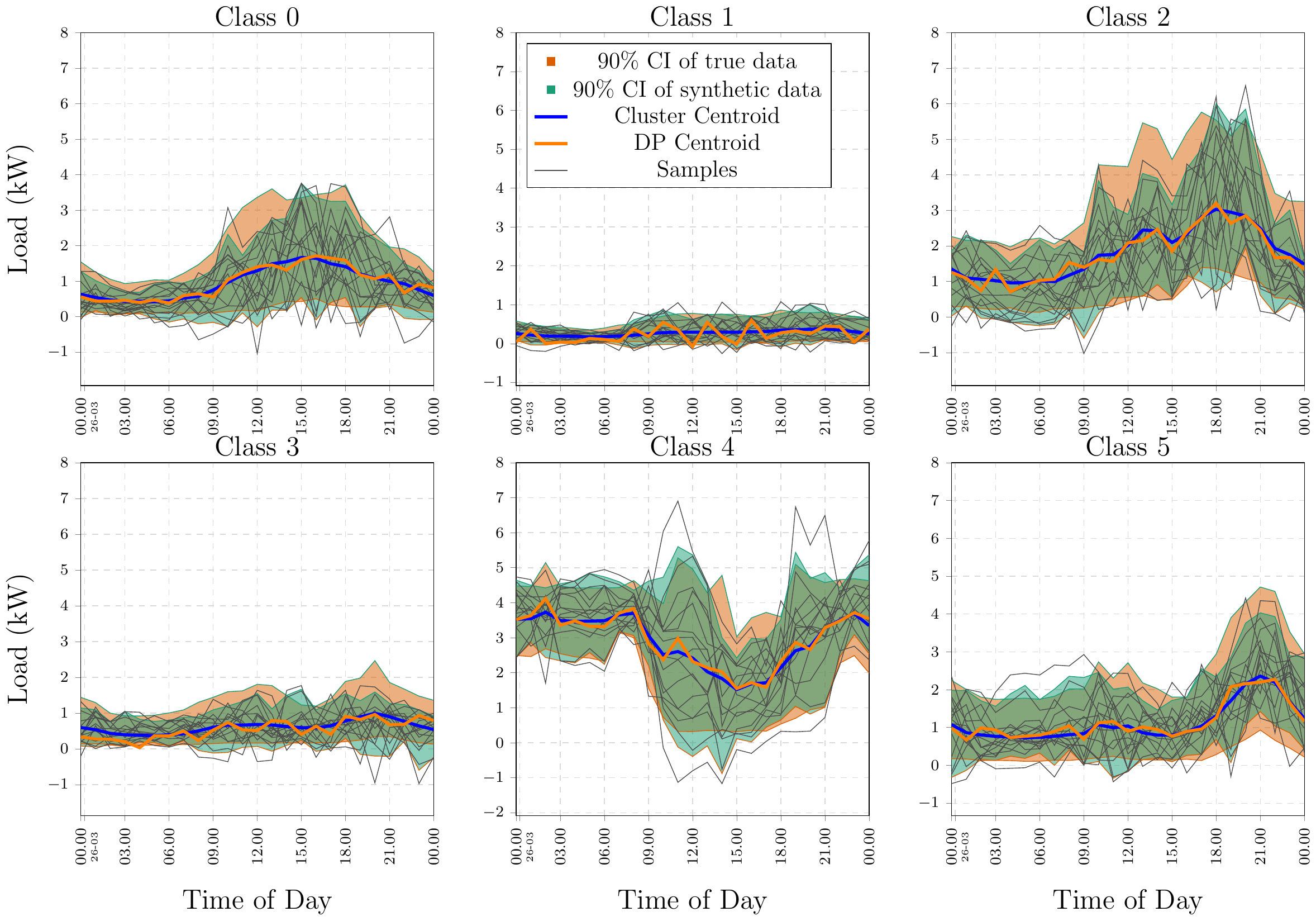}
    \caption{DP Daily Load Shapes of the six clusters without outliers: $\epsilon_c = 30,\delta_c = 0.2$. The brown and green shaded regions indicate the $90\%$ confidence interval (CI) of the true and the synthetically generated data points, respectively. These indicate the region between the $0.05$ and $0.95$ quantiles of the true dataset and that of the fitted log-normal distribution. The synthetic log-normal Daily Load Shapes of the six clusters are shown in gray lines.}
    \label{fig:ami_daily_filtered}
\end{figure*}

\underline{Synthetic Load Profile Generation}:
Now, we use \cref{alg:lognorm} to generate synthetic load profiles for this dataset. In \cref{fig:hists}, we notice that the histograms of samples at each time interval are heavy tailed for every cluster except clusters 2 and 4. Note however that the number of nodes in these two clusters are significantly lower than the number of time intervals. We fit each time interval at each cluster to a group of heavy tailed distribution, and according to the Bayesian Information Criterion (BIC), the log-normal distribution is the best fit with the lowest BIC, for all clusters except $2$ and $4$. Since the number of samples in these two clusters are significantly lower than the number of time intervals (25), fitting any multivariate distribution becomes extremely inaccurate.
Finally, in \cref{fig:ami_daily_filtered}, we show (in gray) $15$ log-normal sample time series for each cluster with $\alpha = 15kW$. Here, the following privacy parameters were utilized: $(\epsilon_c,\delta_c,\epsilon_{\ell},\delta_{\ell}) = (30,0.2,30,0)$.

\begin{figure}[!htbp]
    \centering
    \includegraphics[width=0.5\textwidth]{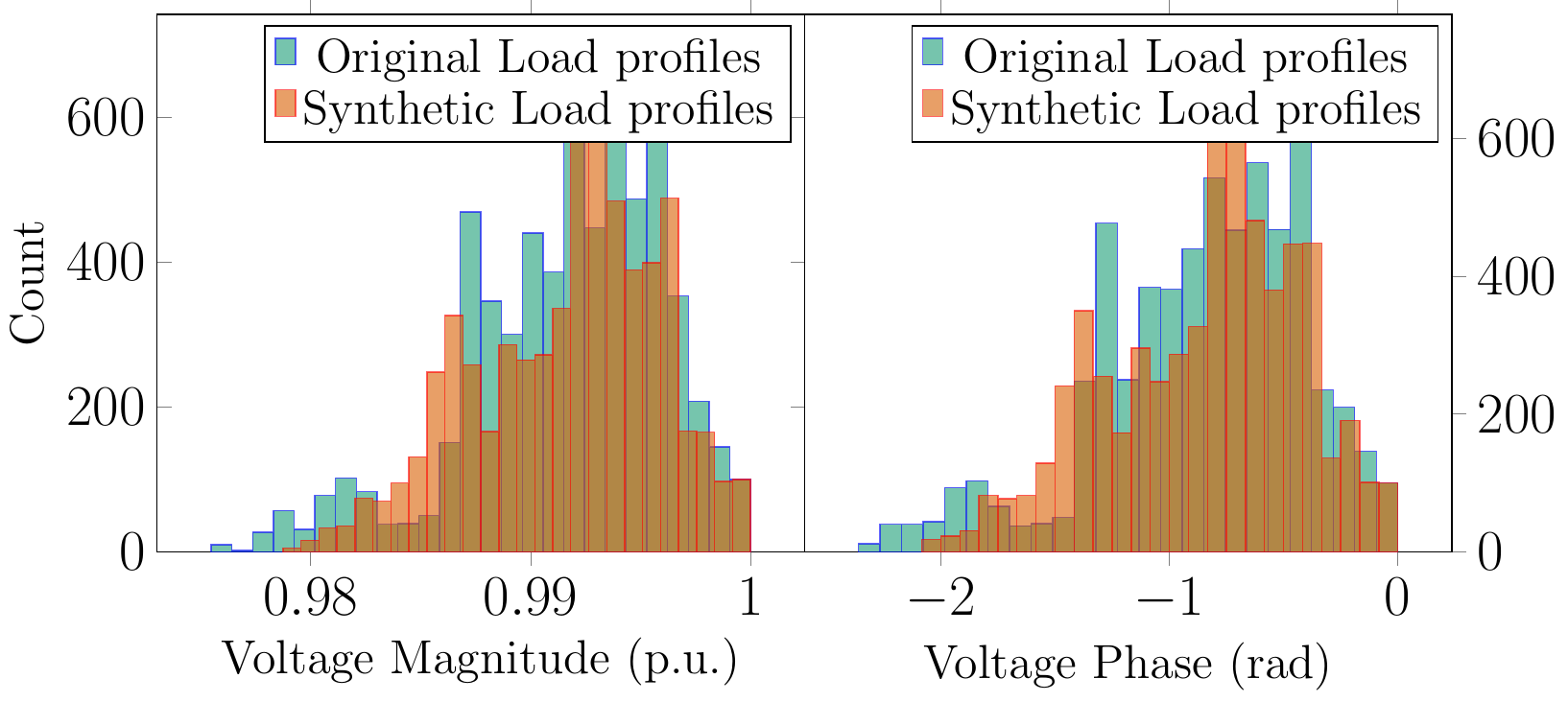}
    \caption{Histogram of the voltage magnitude and phase under true and synthetic load profiles for the IEEE 123-bus case.}
    \label{fig:ieee123}
\end{figure}

\begin{figure}[!htbp]
    \centering
    \includegraphics[width=0.5\textwidth]{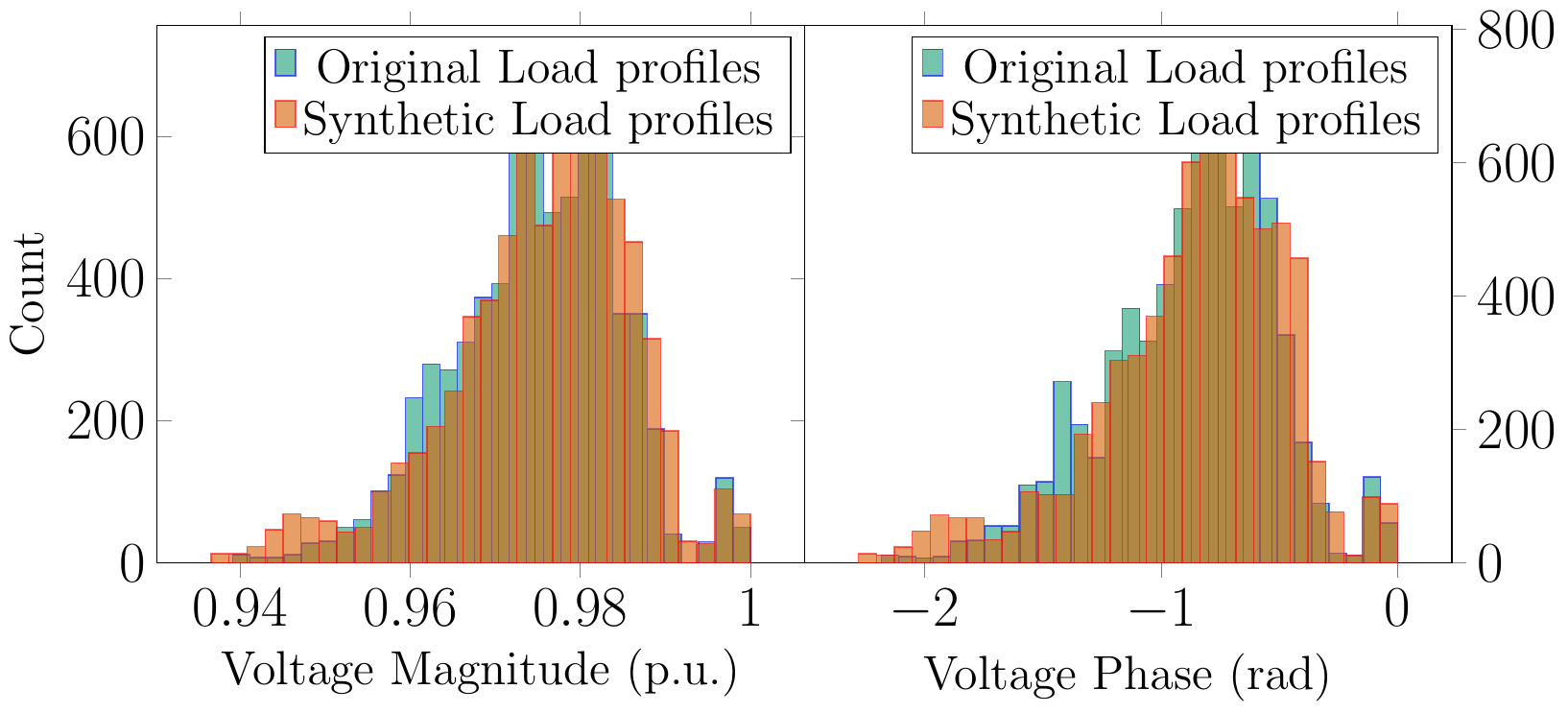}
    \caption{Histogram of the voltage magnitude and phase under true and synthetic load profiles for the MATPOWER 141-bus case.}
    \label{fig:case141}
\end{figure}

\underline{Standard Test Systems}:
The synthetic load profiles generated in the previous section are now tested on two standard test cases, namely the MATPOWER 141-bus radial distribution system from \cite{zimmerman2010matpower} with $P' = 141$ and a modified balanced IEEE-123 test case from \cite{chai2018network} with $P'=123$. The testing methodology is as follows:
\begin{enumerate}[leftmargin=*]
    \item Randomly sample $P'$ houses from the original dataset of $P=1409$ houses, where the six classes are weighed according to their population. Let $m_k$ be the size of class $k$ in the sampled dataset.
    \item Generate $m_k$ samples using \Cref{alg:lognorm} for all classes $k\in[K]$. Now we have $\bm{X}^{(t)} \in \mathbb{R}^{P'\times d}$ consisting of $P'$ true load profiles for $d$ time intervals and a corresponding $\bm{X}^{(s)} \in \mathbb{R}^{P' \times d}$ for the $P'$ synthetic load profiles.
    \item Load a test case with $P'$ buses and set $X^{(t)}_{p,t}$ as the active power load for bus $p$, for all $p \in [P']$. Similarly, the reactive power load is set as $10\%$ of the active power load. 
    \item Run an optimum power flow for this case with the modified loads and collect the voltage magnitude and phase information at each bus.
    \item Repeat steps 3 and 4 for all time intervals $t \in [d]$.
    \item Repeat steps 3, 4, and 5 by using $\bm{X}^{(s)}$ instead of $\bm{X}^{(t)}$.
\end{enumerate}
In \cref{fig:ieee123} and \cref{fig:case141}, we show the histograms of voltage magnitude and phase obtained under both cases with true and synthetic load profiles. The voltage magnitude and phase obtained for the synthetic load profiles provide a good match for those obtained for the true load profiles.

\section{Conclusion}\label{sec:conclusion}
In this paper, we presented an efficient, differentially private mechanism to answer summary statistics about data pertaining to a smart grid. To answer queries about the users in a dataset, such as daily load shapes, we showed the use of clustering and the publication of the extracted clustered centroids rather than publishing individual daily loads. Our algorithm includes a colored Gaussian noise mechanism to guarantee differential privacy about the cluster centroids, and a novel discrete noise mechanism to guarantee differential privacy of the cluster labels. We also demonstrated the utility of our proposed mechanism using numerical simulations by answering queries such as daily load shapes and load duration curves of the houses in a power systems daily load dataset consisting of $1416$ houses. In addition, we showed the importance of domain-specific knowledge to improve the utility of differential privacy. Finally, using the proposed clustering mechanism, we provided a mechanism to generate prototypical daily load shapes for the houses in a dataset.

\appendices

\section{Proof of \texorpdfstring{\Cref{thm:label_DP}}{Theorem 8}}\label[appendix]{app:label_DP}
The joint probability mass function of the noise samples (stacked in a vector $\bm{\nu}$) can be written as:
\begin{equation*}
    f(\bm{\nu}) = (1-\rho)^{\gamma-\|\bm{\nu}\|_0} \left(\frac{\rho}{K-1}\right)^{\|\bm{\nu}\|_0},
\end{equation*}
where $\|\bm{a}\|_0$ is the zero ``norm'' operator that counts the number of non-zero elements in the vector $\bm{a}$. Consider the privacy leakage function:
\begin{equation*}
    L_{\XX}(\tilde{\bm{\ell}}) =  \log \frac{f(\tilde{\mathbb{l}}(X)|X)}{f(\tilde{\mathbb{l}}(X')|X')} = \log \frac{f(\bm{\nu})}{f(\bm{\nu}+\mathbb{l}_{\mathcal{L}}(X')-\mathbb{l}_{\mathcal{L}}(X))},
\end{equation*}
where $\mathbb{l}_\mathcal{L}(X)$ and $\mathbb{l}_\mathcal{L}(X')$ map the points in $X \cap \mathcal{L}$ and $X' \cap \mathcal{L}$, respectively, to their labels. Letting $\mathbb{l}_{\XX} = \mathbb{l}_{\mathcal{L}}(X')-\mathbb{l}_{\mathcal{L}}(X)$, we have:
\begin{align*}
    L_{\XX}(\tilde{\bm{\ell}}) &=  \log \left((\rho^{-1}-1)(K-1)\right)^{\|\bm{\nu}+\mathbb{l}_{\XX}\|_0-\|\bm{\nu}\|_0}\nonumber\\
    &= \left(\|\bm{\nu}+\mathbb{l}_{\XX}\|_0-\|\bm{\nu}\|_0\right)\log \left((\rho^{-1}-1)(K-1)\right).
\end{align*}
For $K\geq 2$ and $\rho < 0.5$, the term $\log \left((\rho^{-1}-1)(K-1)\right)$ is non-negative. Thus, the probability that the absolute value of the privacy leakage function exceeds $\epsilon_\ell$ is:
\begin{align}
    Pr&(L_{\XX}(\tilde{\bm{\ell}}) \geq \epsilon_\ell) = Pr(\|\bm{\nu}+\mathbb{l}_{\XX}\|_0-\|\bm{\nu}\|_0 \geq \ell),\label{eq:label_pl}
\end{align}
where $\ell := \epsilon_{\ell} / \log \left((\rho^{-1}-1)(K-1)\right)$.
\paragraph{Case 1}
When $\ell > \|\mathbb{l}_{\XX}\|_0$, the above probability is $0$. Furthermore, with $\Delta\mathbb{l}~:=~\sup_{X'} \|\mathbb{l}_{\XX}\|_0$, we have:
\begin{align}
    \sup_{X'} Pr(L_{\XX}(\tilde{\bm{\ell}}) \geq \epsilon_\ell) = 0 \quad\text{if}\quad \ell > \Delta\mathbb{l}.\label{eq:label_0}
\end{align}
\paragraph{Case 2}
When $\ell \leq \|\mathbb{l}_{\XX}\|_0$, the calculation of the probability in \cref{eq:label_pl} is not a straightforward task and involves numerical estimation.

Let $M_0$ and $M_c$ be the number of zero elements and the number of elements which satisfy $\nu_p + \mathbb{l}_{\XX,p} = 0$,\footnote{Note that we are using modulo $K$ addition here.} respectively, in $\bm{\nu}$. Based on the values of $\nu_p$ and $\mathbb{l}_{\XX,p}\ne 0$,\footnote{In case of $\mathbb{l}_{\XX,p} = 0$, the value of ($\|\bm{\nu} + \mathbb{l}_{\XX}\|_0 - \|\bm{\nu}\|_0$) does not change for any value of $\nu_p$.} we can make the following observations:
\begin{enumerate}
    \item Whenever $\nu_p = 0$, the value of ($\|\bm{\nu} + \mathbb{l}_{\XX}\|_0 - \|\bm{\nu}\|_0$) increases by $1$.
    \item Whenever $\nu_p \ne 0$ such that $\nu_p + \mathbb{l}_{\XX,p} = 0$, the value of ($\|\bm{\nu} + \mathbb{l}_{\XX}\|_0 - \|\bm{\nu}\|_0$) decreases by $1$.
    \item For the remaining possible value that $\nu_p$ can take, the value of ($\|\bm{\nu} + \mathbb{l}_{\XX}\|_0 - \|\bm{\nu}\|_0$) does not change. The number of such elements in $\bm{\nu}$ is $\|\mathbb{l}_{\XX}\|_0 - M_0 - M_c$.
\end{enumerate}
From their respective definitions, it is clear that $M_0$ and $M_c$ are jointly multinomial, i.e.,
\begin{align}
    &Pr(M_0 = m_0, M_c = m_c) \nonumber\\
    &= \frac{(\|\mathbb{l}_{\XX}\|_0)! (1-\rho)^{m_0} (\frac{\rho}{K-1})^{m_c}(\frac{\rho(K-2)}{K-1})^{(\|\mathbb{l}_{\XX}\|_0 - m_0 - m_c)}}{m_0! m_c! (\|\mathbb{l}_{\XX}\|_0 - m_0 - m_c)!},
\end{align}
and it can also be shown that $\|\bm{\nu}+\mathbb{l}_{\XX}\|_0-\|\bm{\nu}\|_0 = M_0 - M_c$. When $\ell \leq \|\mathbb{l}_{\XX}\|_0$, \cref{eq:label_pl} can be simplified to:
\begin{align}
    &Pr(\|\bm{\nu}+\mathbb{l}_{\XX}\|_0-\|\bm{\nu}\|_0 \geq \ell) \nonumber\\
    &= \sum_{m_c = 0}^{\|\mathbb{l}_{\XX}\|_0}\sum_{m_0 = \ell + m_c}^{\|\mathbb{l}_{\XX}\|_0} Pr(M_0 = m_0, M_c = m_c)\nonumber\\
    &= (\|\mathbb{l}_{\XX}\|_0)! \left[ \frac{\rho(K-2)}{K-1}\right]^{\|\mathbb{l}_{\XX}\|_0} \times \nonumber\\
    & \sum_{m_c = 0}^{\|\mathbb{l}_{\XX}\|_0}\sum_{m_0 = \ell + m_c}^{\|\mathbb{l}_{\XX}\|_0} \frac{\left[ \frac{(K-1)(1-\rho)}{\rho} \right]^{m_0} (K-2) ^{-(m_0+m_c)}}{m_0! m_c! (\|\mathbb{l}_{\XX}\|_0 - m_0 - m_c)!}\label{eq:label_color_delta}
\end{align}
Since the expression \cref{eq:label_color_delta} grows monotonically with $\|\mathbb{l}_{\XX}\|_0$, we can define the sensitivity as $\Delta\mathbb{l} := \sup_{X'} \|\mathbb{l}_{\XX}\|_0$. Thus, 
\begin{align}
    &\sup_{X'} Pr(L_{\XX}(\tilde{\bm{\ell}}) \geq \epsilon_\ell) = \Delta\mathbb{l}! \left[ \frac{\rho(K-2)}{K-1}\right]^{\Delta\mathbb{l}} \times \nonumber\\
    & \sum_{m_c = 0}^{\Delta\mathbb{l}}\sum_{m_0 = \ell + m_c}^{\Delta\mathbb{l}} \frac{\left[ \frac{(K-1)(1-\rho)}{\rho} \right]^{m_0} (K-2) ^{-(m_0+m_c)}}{m_0! m_c! (\Delta\mathbb{l} - m_0 - m_c)!}\label{eq:label_n0}
\end{align}
From \cref{eq:label_0} and \cref{eq:label_n0}, the proposed mechanism is $(\epsilon_\ell,\delta_\ell)-$PDP and thus, it is also $(\epsilon_\ell,\delta_\ell)-$DP from \cref{thm:PDP-DP}.

\ifCLASSOPTIONcaptionsoff
  \newpage
\fi


\bibliographystyle{IEEEtran}
\bibliography{ref}

\begin{thebibliography}{10}
\providecommand{\url}[1]{#1}
\csname url@samestyle\endcsname
\providecommand{\newblock}{\relax}
\providecommand{\bibinfo}[2]{#2}
\providecommand{\BIBentrySTDinterwordspacing}{\spaceskip=0pt\relax}
\providecommand{\BIBentryALTinterwordstretchfactor}{4}
\providecommand{\BIBentryALTinterwordspacing}{\spaceskip=\fontdimen2\font plus
\BIBentryALTinterwordstretchfactor\fontdimen3\font minus
  \fontdimen4\font\relax}
\providecommand{\BIBforeignlanguage}[2]{{%
\expandafter\ifx\csname l@#1\endcsname\relax
\typeout{** WARNING: IEEEtran.bst: No hyphenation pattern has been}%
\typeout{** loaded for the language `#1'. Using the pattern for}%
\typeout{** the default language instead.}%
\else
\language=\csname l@#1\endcsname
\fi
#2}}
\providecommand{\BIBdecl}{\relax}
\BIBdecl

\bibitem{diamantoulakis2015big}
P.~D. Diamantoulakis, V.~M. Kapinas, and G.~K. Karagiannidis, ``{Big Data
  Analytics for Dynamic Energy Management in Smart Grids},'' \emph{Big Data
  Research}, vol.~2, no.~3, pp. 94--101, 2015.

\bibitem{zhang2018big}
Y.~Zhang, T.~Huang, and E.~F. Bompard, ``{Big Data Analytics in Smart Grids: A
  Review},'' \emph{Energy informatics}, vol.~1, no.~1, pp. 1--24, 2018.

\bibitem{wang2018review}
Y.~Wang, Q.~Chen, T.~Hong, and C.~Kang, ``{Review of Smart Meter Data
  Analytics: Applications, Methodologies, and Challenges},'' \emph{IEEE
  Transactions on Smart Grid}, vol.~10, no.~3, pp. 3125--3148, 2018.

\bibitem{yang2013review}
S.-l. Yang, C.~Shen \emph{et~al.}, ``A review of electric load classification
  in smart grid environment,'' \emph{Renewable and Sustainable Energy Reviews},
  vol.~24, pp. 103--110, 2013.

\bibitem{ramos2012typical}
S.~Ramos, J.~Duarte, J.~Soares, Z.~Vale, and F.~J. Duarte, ``{Typical Load
  Profiles in the Smart Grid Context -- A Clustering Methods Comparison},'' in
  \emph{2012 IEEE Power and Energy Society General Meeting}.\hskip 1em plus
  0.5em minus 0.4em\relax IEEE, 2012, pp. 1--8.

\bibitem{sharma2014electrical}
D.~D. Sharma and S.~Singh, ``{Electrical Load Profile Analysis and Peak Load
  Assessment Using Clustering Technique},'' in \emph{2014 IEEE PES General
  Meeting Conference \& Exposition}, 2014.

\bibitem{asghar2017smart}
M.~R. Asghar, G.~D{\'a}n, D.~Miorandi, and I.~Chlamtac, ``{Smart Meter Data
  Privacy: A Survey},'' \emph{IEEE Communications Surveys \& Tutorials},
  vol.~19, no.~4, pp. 2820--2835, 2017.

\bibitem{ruj2013decentralized}
S.~Ruj and A.~Nayak, ``{A Decentralized Security Framework for Data Aggregation
  and Access Control in Smart Grids},'' \emph{IEEE transactions on smart grid},
  vol.~4, no.~1, pp. 196--205, 2013.

\bibitem{efthymiou2010smart}
C.~Efthymiou and G.~Kalogridis, ``{Smart Grid Privacy via Anonymization of
  Smart Metering Data},'' in \emph{2010 first IEEE international conference on
  smart grid communications}.\hskip 1em plus 0.5em minus 0.4em\relax IEEE,
  2010, pp. 238--243.

\bibitem{Narayanan2014No-silver-bulle}
A.~Narayanan and E.~W. Felten, ``{No Silver Bullet: De-identification Still
  Doesn't Work},''
  \url{http://randomwalker.info/publications/no-silver-bullet-de-identification.pdf},
  July 9, 2014.

\bibitem{barbaro2006face}
M.~Barbaro, T.~Zeller, and S.~Hansell, ``{A face is exposed for AOL searcher
  no. 4417749},'' \emph{New York Times}, vol.~9, no. 2008, p.~8, 2006.

\bibitem{Narayanan2008-short}
A.~Narayanan and V.~Shmatikov, ``{Robust De-Anonymization of Large Sparse
  Datasets},'' in \emph{29th IEEE Symposium on Security and Privacy}, May 2008.

\bibitem{sweeney2013identifying}
L.~Sweeney, A.~Abu, and J.~Winn, ``{Identifying Participants in the Personal
  Genome Project by Name},'' \emph{Available at SSRN 2257732}, 2013.

\bibitem{15_15rule}
P.~U.~C. of~the State~of Colorado, ``{Decision No. R11-0922},'' \emph{Proposed
  Rules Relating to Smart Grid Data Privacy for Electric Utilities}, 2011.

\bibitem{dwork2006calibrating}
C.~Dwork, F.~McSherry, K.~Nissim, and A.~Smith, ``{Calibrating Noise to
  Sensitivity in Private Data Analysis},'' in \emph{Theory of cryptography
  conference}, 2006, pp. 265--284.

\bibitem{yu2016outlier}
Q.~Yu, Y.~Luo, C.~Chen, and X.~Ding, ``{Outlier-Eliminated K-Means Clustering
  Algorithm Based on Differential Privacy Preservation},'' \emph{Applied
  Intelligence}, vol.~45, no.~4, pp. 1179--1191, 2016.

\bibitem{balcan2017differentially}
M.-F. Balcan, T.~Dick, Y.~Liang, W.~Mou, and H.~Zhang, ``{Differentially
  Private Clustering in High-Dimensional Euclidean Spaces},'' in
  \emph{International Conference on Machine Learning}.\hskip 1em plus 0.5em
  minus 0.4em\relax PMLR, 2017, pp. 322--331.

\bibitem{ren2017dplk}
J.~Ren, J.~Xiong, Z.~Yao, R.~Ma, and M.~Lin, ``Dplk-means: A novel differential
  privacy k-means mechanism,'' in \emph{2017 IEEE Second International
  Conference on Data Science in Cyberspace (DSC)}.\hskip 1em plus 0.5em minus
  0.4em\relax IEEE, 2017, pp. 133--139.

\bibitem{xia2020distributed}
C.~Xia, J.~Hua, W.~Tong, and S.~Zhong, ``{Distributed K-Means Clustering
  Guaranteeing Local Differential Privacy},'' \emph{Computers \& Security},
  vol.~90, p. 101699, 2020.

\bibitem{lu2020differentially}
Z.~Lu and H.~Shen, ``{Differentially Private K-Means Clustering With Guaranteed
  Convergence},'' \emph{arXiv preprint arXiv:2002.01043}, 2020.

\bibitem{ni2021utility}
T.~Ni, M.~Qiao, Z.~Chen, S.~Zhang, and H.~Zhong, ``{Utility-Efficient
  Differentially Private K-Means Clustering Based on Cluster Merging},''
  \emph{Neurocomputing}, vol. 424, pp. 205--214, 2021.

\bibitem{pinceti2019data}
A.~Pinceti, O.~Kosut, and L.~Sankar, ``{Data-Driven Generation of Synthetic
  Load Datasets Preserving Spatio-Temporal Features},'' in \emph{2019 IEEE
  Power \& Energy Society General Meeting (PESGM)}, 2019.

\bibitem{el2020data}
S.~El~Kababji and P.~Srikantha, ``{A Data-Driven Approach for Generating
  Synthetic Load Patterns and Usage Habits},'' \emph{IEEE Transactions on Smart
  Grid}, vol.~11, no.~6, pp. 4984--4995, 2020.

\bibitem{fekri2020generating}
M.~N. Fekri, A.~M. Ghosh, and K.~Grolinger, ``{Generating Energy Data for
  Machine Learning With Recurrent Generative Adversarial Networks},''
  \emph{Energies}, vol.~13, no.~1, p. 130, 2020.

\bibitem{pinceti2021synthetic}
A.~Pinceti, L.~Sankar, and O.~Kosut, ``{Synthetic Time-Series Load Data via
  Conditional Generative Adversarial Networks},'' \emph{arXiv preprint
  arXiv:2107.03545}, 2021.

\bibitem{ravi2021colored}
N.~Ravi, A.~Scaglione, and S.~Peisert, ``{Colored Noise Mechanism for
  Differentially Private Clustering},'' \emph{arXiv preprint arXiv:2111.07850},
  2021.

\bibitem{lou2017cost}
X.~Lou, R.~Tan, D.~K. Yau, and P.~Cheng, ``{Cost of Differential Privacy in
  Demand Reporting for Smart Grid Economic Dispatch},'' in \emph{IEEE INFOCOM
  2017-IEEE Conference on Computer Communications}.\hskip 1em plus 0.5em minus
  0.4em\relax IEEE, 2017, pp. 1--9.

\bibitem{dwork2006our}
C.~Dwork, K.~Kenthapadi, F.~McSherry, I.~Mironov, and M.~Naor, ``{Our Data,
  Ourselves: Privacy Via Distributed Noise Generation},'' in \emph{Annual
  International Conference on the Theory and Applications of Cryptographic
  Techniques}.\hskip 1em plus 0.5em minus 0.4em\relax Springer, 2006, pp.
  486--503.

\bibitem{machanavajjhala2008privacy}
A.~Machanavajjhala, D.~Kifer, J.~Abowd, J.~Gehrke, and L.~Vilhuber, ``Privacy:
  Theory meets practice on the map,'' in \emph{2008 IEEE 24th international
  conference on data engineering}.\hskip 1em plus 0.5em minus 0.4em\relax IEEE,
  2008, pp. 277--286.

\bibitem{mcclure2015relaxations}
D.~McClure, ``{Relaxations of Differential Privacy and Risk/Utility Evaluations
  of Synthetic Data and Fidelity Measures},'' Ph.D. dissertation, Duke
  University, 2015.

\bibitem{dwork2014algorithmic}
C.~Dwork, A.~Roth \emph{et~al.}, ``{The Algorithmic Foundations of Differential
  Privacy},'' \emph{Found. Trends Theor. Comput. Sci.}, vol.~9, no. 3-4, pp.
  211--407, 2014.

\bibitem{press1990savitzky}
W.~H. Press and S.~A. Teukolsky, ``Savitzky-golay smoothing filters,''
  \emph{Computers in Physics}, vol.~4, no.~6, pp. 669--672, 1990.

\bibitem{jiang2016wishart}
W.~Jiang, C.~Xie, and Z.~Zhang, ``{Wishart Mechanism for Differentially Private
  Principal Components Analysis},'' in \emph{Proc. AAAI Conference on
  Artificial Intelligence}, vol.~30, no.~1, 2016.

\bibitem{mead1992review}
A.~Mead, ``{Review of the Development of Multidimensional Scaling Methods},''
  \emph{Journal of the Royal Statistical Society: Series D (The Statistician)},
  vol.~41, no.~1, pp. 27--39, 1992.

\bibitem{zimmerman2010matpower}
R.~D. Zimmerman, C.~E. Murillo-S{\'a}nchez, and R.~J. Thomas, ``Matpower:
  Steady-state operations, planning, and analysis tools for power systems
  research and education,'' \emph{IEEE Transactions on power systems}, vol.~26,
  no.~1, pp. 12--19, 2010.

\bibitem{chai2018network}
Y.~Chai, L.~Guo, C.~Wang, Z.~Zhao, X.~Du, and J.~Pan, ``Network partition and
  voltage coordination control for distribution networks with high penetration
  of distributed pv units,'' \emph{IEEE Transactions on Power Systems},
  vol.~33, no.~3, pp. 3396--3407, 2018.

\end{thebibliography}

\end{document}